\documentclass[preprint,floatfix]{revtex4}

\usepackage{graphicx} 
\usepackage{dcolumn} 
\usepackage{amssymb}
\usepackage{amsmath}
\usepackage{bm}
\usepackage[dvips]{color}
\usepackage{times}
\usepackage{graphicx}
\usepackage{ulem}
\usepackage[justification=raggedright]{caption}

\usepackage{subcaption}
\usepackage{tabularx}
\usepackage{appendix}

\usepackage{amstext}
\usepackage{array}

\usepackage{bbm}

\usepackage[colorlinks=true
,urlcolor=blue
,anchorcolor=blue
,citecolor=red
,filecolor=blue
,linkcolor=blue
,menucolor=blue
,pagecolor=blue
,linktocpage=true
,pdfproducer=medialab
]{hyperref}

\allowdisplaybreaks 

\newcommand{\ncmd}{\newcommand}
\ncmd{\lt}{\left}
\ncmd{\rt}{\right}
\ncmd{\tr}[1]{\mbox{Tr}\lt[{#1}\rt]}
\ncmd{\half}{\frac{1}{2}}
\ncmd{\eps}{\epsilon}
\ncmd{\veps}{\varepsilon}
\ncmd{\dgr}{\dagger}
\ncmd{\sig}{\sigma}
\ncmd{\rtarw}{\rightarrow}
\ncmd{\Rt}{\Rightarrow}
\ncmd{\abs}[1]{\lt|{#1}\rt|}
\ncmd{\avg}[1]{\lt<{#1}\rt>}
\ncmd{\dl}{\delta}
\ncmd{\Dl}{\Delta}
\ncmd{\sgn}[1]{\mbox{sgn}\lt(#1\rt)}
\ncmd{\kap}{\kappa}
\ncmd{\wtil}[1]{\widetilde{#1}}
\ncmd{\thrfr}{\therefore}
\ncmd{\eq}[1]{Eq. \eqref{#1}}
\ncmd{\fig}[1]{Fig. \ref{#1}}
\ncmd{\Lam}{\Lambda}
\ncmd{\lam}{\lambda}
\ncmd{\dow}{\partial}
\ncmd{\ordr}[1]{\mathcal{O}\lt(#1\rt)}
\ncmd{\dsty}{\displaystyle}
\ncmd{\alert}[1]{\color{red}{#1}}
\ncmd{\mc}{\mathcal}
\ncmd{\mbf}[1]{\mathbf{#1}}
\ncmd{\Deriv}[2]{\frac{d{#1}}{d{#2}}}
\ncmd{\ParDeriv}[2]{\frac{\partial{#1}}{\partial{#2}}}
\ncmd{\step}[1]{\Theta\lt(#1\rt)}
\ncmd{\td}{\tilde} 
\ncmd{\what}{\widehat}
\ncmd{\om}{\omega}
\ncmd{\Om}{\Omega}
\ncmd{\vrho}{\varrho}
\ncmd{\vsig}{\varsigma}
\ncmd{\vkap}{\varkappa}
\ncmd{\bqa}{\begin{eqnarray}} 
\ncmd{\eqa}{\end{eqnarray}}
\ncmd{\nn}{\nonumber \\}
\ncmd{\nnum}{\nonumber}
\ncmd{\comment}[1]{{\color{red}{#1}}}
\definecolor{new_color}{RGB}{50,155,0}

\begin{document}

\title{
Quasi-Local Strange Metal 
}

\author{Shouvik Sur$^{1}$ and Sung-Sik Lee$^{1,2}$\\
\vspace{0.3cm}
{\normalsize{$^1$Department of Physics $\&$ Astronomy, 
McMaster University,}}\\
{\normalsize{1280 Main St. W., Hamilton ON L8S 4M1, Canada}}
\vspace{0.2cm}\\
{\normalsize{$^2$Perimeter Institute for Theoretical 
Physics,}}\\
{\normalsize{31 Caroline St. N., Waterloo ON N2L 2Y5, 
Canada}}
}

\date{\today}

\begin{abstract}

One of the key factors that determine 
the fates of quantum many-body systems
in the zero temperature limit
is the competition between 
kinetic energy that delocalizes particles in space 
and interaction that promotes localization.
While one dominates over the other in conventional metals and insulators,
exotic states can arise at quantum critical points
where none of them clearly wins.
Here we present a novel metallic state 
that emerges at
an antiferromagnetic (AF) quantum critical point 
in the presence of one-dimensional Fermi surfaces
embedded in space dimensions three and below.
At the critical point,
interactions between particles are 
screened to zero in the low energy limit
at the same time
the kinetic energy is suppressed in certain spatial directions
to the leading order in a perturbative expansion that
becomes asymptotically exact in three dimensions.
The resulting \textit{dispersionless} and \textit{interactionless} state exhibits
distinct quasi-local strange metallic behaviors 
due to a subtle dynamical balance between 
screening and infrared singularity
caused by spontaneous reduction of effective dimensionality.
The strange metal, which is stable near three dimensions, shows enhanced fluctuations of bond density waves, d-wave pairing, and pair density waves.

\end{abstract}

\maketitle

The richness of exotic zero-temperature states 
in condensed matter systems\cite{wen2004quantum,sachdev2001quantum}
can be attributed to quantum fluctuations 
driven by kinetic energy and interaction 
which can not be simultaneously minimized
due to the uncertainty principle.
In conventional metals, 
kinetic energy plays the dominant role,
and interactions only dress electrons into quasiparticles 
which survive as coherent excitations in the absence of instabilities\cite{RevModPhys.66.129,1992hep.th...10046P}.
The existence of well defined quasiparticle excitations
is the cornerstone of Landau Fermi liquid theory\cite{LFL},
which successfully explains a large class of metals.
However, the Fermi liquid theory breaks down 
at the verge of spontaneous formation of order in metals\cite{PhysRevB.14.1165,PhysRevB.48.7183,RevModPhys.79.1015}.
Near continuous quantum phase transitions,
new metallic states can arise
as quantum fluctuations of order parameter
destroy the coherence of quasiparticles 
through interactions 
that persist down to the zero energy limit\cite{RevModPhys.73.797,PhysRevB.78.035103}.
Systematic understanding 
of the resulting strange metallic states is still lacking, 
although there exist some examples 
whose universal behaviors in the low energy limit
can be understood within controlled theoretical frameworks\cite{nayak1994non,PhysRevB.82.045121,jiang2013non,PhysRevB.88.245106,2013arXiv1310.7543S}.

Antiferromagnetic (AF) quantum phase transition commonly 
arises in strongly correlated systems including
electron doped cuprates\cite{PhysRevLett.105.247002}, 
iron pnictides\cite{Hashimoto22062012} and 
heavy fermion compounds\cite{park2006hidden}.
In two space dimensions,
it has been shown that 
the interaction between the AF mode and itinerant electrons
qualitatively modify the dynamics of the system
at the critical point\cite{PhysRevLett.84.5608,PhysRevLett.93.255702}.
A recent numerical simulation shows 
a strong enhancement of superconducting correlations
near the AF critical point\cite{Berg21122012}.
However, the precise nature  
of the putative strange metallic state 
has not been understood yet
due to a lack of theoretical control over
the strongly coupled theory
that governs the critical point\cite{PhysRevB.82.075128}.
In this article, based on a controlled expansion,
we show that a novel quantum state arises
at the AF quantum critical point 
in metals that support one-dimensional Fermi surface
through a non-trivial interplay between 
kinetic energy and interactions.
To the lowest order in the perturbative expansion 
that becomes asymptotically exact at low energies in three dimensions,
we find that quasiparticles are destroyed 
even though the interaction between electrons 
and the AF mode is screened to zero in the low energy limit.
This unusual behavior is possible as
the system develops an infinite sensitivity to the interaction
through the kinetic energies that become dispersionless in certain spatial directions.
The dynamical balance 
between vanishing kinetic energy and interactions
results in a stable {\it quasi-local strange metal} 
which supports incoherent single-particle excitations
and enhanced correlations for various competing orders.

\begin{figure}[!t]
\centering
 \includegraphics[scale=0.5]{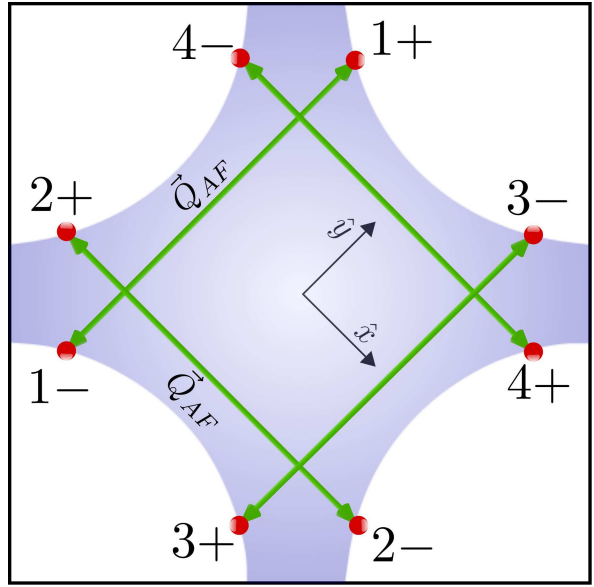}
 \caption{
A two-dimensional Fermi surface
where the shaded (unshaded) region represents
occupied (unoccupied) states in momentum space.
The hot spots on the Fermi surface are denoted as (red) dots.
The (green) arrows represent the AF wavevector $\vec{Q}_{AF}$. 
}
\label{fig:hot_spots}
\end{figure}

{\bf Model and dimensional regularization}.
We first consider two space dimensions.
Although the specific lattice is not crucial for the 
following discussion,
we consider the square lattice 
with the nearest and next-nearest neighbor hoppings.
For electron density close to half-filling (one electron per site),
the system supports a Fermi surface 
shown in Fig. \ref{fig:hot_spots}.
The minimal theory that describes the AF critical point 
in the two dimensional metal
includes the collective AF fluctuations that are coupled to electrons near the hot spots, which are the set of points on the Fermi surface connected by the AF wavevector  \cite{PhysRevLett.84.5608,PhysRevLett.93.255702,PhysRevB.82.075128}. 
In this paper we consider the collinear AF order with a commensurate wavevector that is denoted as arrows in Fig. \ref{fig:hot_spots}.
If the AF order is incommensurate or non-collinear,  
the critical theory is modified from the one for the collinear AF order with a commensurate wavevector.
As will be shown later,  the simplest case we consider here already has quite intricate structures.

The action for the commensurate AF mode and the electrons near the hot spots reads
\begin{align}
  \mathcal{S} &= 
\sum_{n=1}^4 \sum_{m=\pm} \sum_{\sigma=\uparrow,\downarrow} 
\int \frac{d^3k}{(2\pi)^3} ~ 
{\psi}^{(m)*}_{n,\sigma}(k)
\left[ ik_0 ~ + e^{m}_n(\vec k)  \right] 
\psi^{(m)}_{n,\sigma}(k) \nn
& \quad + \half \int \frac{d^3q}{(2\pi)^3} 
\left[ q_0^2 + c^2 | \vec q |^2 \right] \vec{\phi}(-q) 
\cdot \vec{\phi}(q) \nn
  & \quad + g_0 \sum_{n=1}^4 
\sum_{\sigma,\sigma'=\uparrow,\downarrow}
 \int \frac{d^3k}{(2\pi)^3} \frac{d^3q}{(2\pi)^3} ~ 
\Bigl[
\vec{\phi}(q) \cdot 
{\psi}^{(+)*}_{n,\sigma}(k+q) \vec{\tau}_{\sigma,\sigma'}  
\psi^{(-)}_{n,\sigma'} (k) 
+ c.c. \Bigr] \nn
& \quad + \frac{u_0}{4!} 
\int \frac{d^3k_1}{(2\pi)^3}
\frac{d^3k_2}{(2\pi)^3} 
\frac{d^3q}{(2\pi)^3}
 \lt[ \vec{\phi}(k_1+q) \cdot \vec{\phi}(k_2-q)\rt] 
\lt[ \vec{\phi}(k_1) \cdot \vec{\phi}(k_2) \rt]. 
\label{eq:3D_theory}
\end{align}
Here $k=(k_0,\vec k)$ denotes 
frequency and two-dimensional momentum $\vec k =(k_x,k_y)$.
$\psi^{(\pm)}_{n,\sigma}$'s are fermionic fields 
which represent electrons of spin $\sigma=\uparrow, \downarrow$
near the hot spots labeled by $n=1,2,3,4$, $m=\pm$,
as is shown in Fig. \ref{fig:hot_spots}.
The axis in momentum space has been chosen 
such that the AF wavevector becomes 
${\vec Q}_{AF} = \pm  \sqrt{2} \pi \hat y = \pm \sqrt{2} \pi \hat x$
up to the reciprocal vectors $\sqrt{2} \pi ( \hat x \pm \hat y)$.
In this coordinate, the energy dispersions 
of the fermions near the hot spots can be written as
$e_{1}^{\pm}(\vec k) = -e_{3}^{\pm}(\vec k) =   v k_x \pm k_y$,
$e_{2}^{\pm}(\vec k) = -e_{4}^{\pm}(\vec k) = \mp k_x + v k_y$,
where $\vec k$ represents deviation of momentum away 
from each hot spot.
It is noted that local curvature of the Fermi surface
can be ignored because the $k$-linear terms dominate
at low energies.
The component of Fermi velocity parallel to ${\vec Q}_{AF}$ at  
each hot spot is set to be unity up to sign by rescaling
$\vec k$.
$v$ measures the component of the 
Fermi velocity perpendicular to ${\vec Q}_{AF}$.
If $v$ was zero, the hot spots connected by $\vec Q_{AF}$
would be perfectly nested.
$\vec{\phi}(q)$ represents three components of boson field 
which describes the fluctuating AF order parameter 
carrying frequency $q_0$ and momentum ${\vec Q}_{AF} + \vec q$.
$\vec \tau$ represents the three generators of the $SU(2)$ group.
$c$ is the velocity of the AF collective mode.
$g_0$ is the Yukawa coupling between the collective mode
and the electrons near the hot spots,
and $u_0$ is the  quartic 
coupling between the collective modes.
$v,c,g_0,u_0$ are genuine parameters of the theory,
which can not be removed by 
redefinition of momentum or fields.

In two dimensions, 
the perturbative expansion in $g_0$, $u_0$ fails 
because the couplings grow rapidly 
as the length scale is increased
under the renormalization group (RG) flow.
Although the growth of the couplings is tamed by screening,
it is hard to follow the RG flow 
because the flow will stop (if it does) 
outside the perturbative window.
In higher dimensions, the growth of the couplings becomes slower. 
Therefore we aim to tune the space dimension such that
the balance between the slow growth of the couplings and the screening
stabilizes the interacting theory at weak coupling.
Here we increase the co-dimension of the Fermi surface
while fixing its dimension
to be one.
A mere increase of co-dimension of the Fermi surface
introduces a non-locality in the kinetic energy\cite{PhysRevLett.102.046406}.
In order to keep locality of the theory,
we introduce two-component spinors\cite{PhysRevB.88.245106},
by combining fermion fields on opposite sides of the Fermi surface,
$\Psi_{1,\sigma}=(\psi^{(+)}_{1,\sigma},\psi^{(+)}_{3,\sigma})^T$,
$\Psi_{2,\sigma}=(\psi^{(+)}_{2,\sigma},\psi^{(+)}_{4,\sigma})^T$,
$\Psi_{3,\sigma}=(\psi^{(-)}_{1,\sigma}, - \psi^{(-)}_{3,\sigma})^T$,
$\Psi_{4,\sigma}=(\psi^{(-)}_{2,\sigma}, - \psi^{(-)}_{4,\sigma})^T$
and writing the kinetic term of the fermions as
$S_0 = \sum_{n=1}^4 
\sum_{\sigma=\uparrow,\downarrow} 
\int \frac{d^{3}{k}}{(2\pi)^{3}} ~
\bar{\Psi}_{n,\sigma}(k) 
\Bigl[ i \gamma_0 k_0 + i \gamma_{1} \varepsilon_n(\vec{k}) \Bigr] 
\Psi_{n,\sigma}(k)$,
where $\gamma_0 = \sigma_y, \gamma_1 = \sigma_x$,
$\bar \Psi_{n,\sigma}= \Psi^\dagger_{n,\sigma} \gamma_0$
with 
$\varepsilon_1(\vec k) = e_{1}^{+}(\vec k)$,
$\varepsilon_2(\vec k) = e_{2}^{+}(\vec k)$,
$\varepsilon_3(\vec k) = e_{1}^{-}(\vec k)$,
$\varepsilon_4(\vec k) = e_{2}^{-}(\vec k)$.
Now we add $(d-2)$ extra dimensions 
which are perpendicular to the Fermi surface.
We also generalize the $SU(2)$ group to $SU(N_c)$,
and introduce $N_f$ flavors of fermion
to write a general theory, 
\begin{align}
  \mathcal{S} &= 
\sum_{n=1}^4 
\sum_{\sigma=1}^{N_c} 
\sum_{j=1}^{N_f} 
\int 
dk~
\bar{\Psi}_{n,\sigma,j}(k) 
\Bigl[ i\mathbf{\Gamma} \cdot \mathbf{K}  
+ i \gamma_{d-1} \varepsilon_n(\vec{k}) \Bigr] 
\Psi_{n,\sigma,j}(k)
\nn 
&  + \frac{1}{4} 
\int dq~
\Bigl[ \abs{\mathbf{Q}}^2 + c^2 | \vec q |^2 \Bigr]
\tr { \Phi(-q)  ~ \Phi(q) } \nn
  & + i \frac{g \mu^{(3-d)/2}}{\sqrt{N_f}}
\sum_{n=1}^4
\sum_{\sigma,\sigma'=1}^{N_c}
\sum_{j=1}^{N_f} 
\int 
dk dq ~
\Bigl[
\bar{\Psi}_{\bar n,\sigma,j}(k+q) 
\Phi_{\sigma,\sigma'}(q)
\gamma_{d-1} 
\Psi_{n,\sigma',j} (k) 
\Bigr]  \nn
&  + \frac{\mu^{3-d} }{4}
\int 
dk_1 dk_2 dq ~
 \Bigl[ 
  u_1 \tr { \Phi(k_1+q)  \Phi(k_2-q) } \tr { \Phi(k_1)  \Phi(k_2) } \nn
&~~~~~~~~~~~~~~~~~~~~~~~~~~~~~~~~ + u_2 \tr { \Phi(k_1+q)  \Phi(k_2-q)  \Phi(k_1)  \Phi(k_2) }
\Bigr]. 
\label{eq: generalD_theory}
\end{align}
Here 
$dk \equiv \frac{d^{d+1}k}{(2\pi)^{d+1}}$ and
$k=({\bf K}, \vec k)$ is $(d+1)$-dimensional vector.
$\vec k = (k_x, k_y)$ represents the original two-dimensional momentum
and ${\bf K}=(k_0,k_1,\ldots,k_{d-2})$ includes frequency
and momentum components along the $(d-2)$ new directions
present in $d>2$. 
$({\bf \Gamma}, \gamma_{d-1})$ with ${\bf 
\Gamma}=(\gamma_0,\gamma_1,\ldots,\gamma_{d-2})$
represent $(d-1)$-dimensional gamma matrices that satisfy
the Clifford algebra, 
$\{ \gamma_\mu, \gamma_\nu \} = 2 I \delta_{\mu \nu}$ with 
$\tr I = 2$.
$\Psi_{n,\sigma,j}$ with 
$\sigma=1,2,\ldots,N_c$ and $j=1,2,\ldots,N_f$ 
is in the fundamental representation of 
$SU(N_c)$ spin group and $SU(N_f)$ flavor group.
$\Phi(q) = \sum_{a=1}^{N_c^2-1} \phi^a(q) \tau^a$ is a matrix field
where $\tau^a$'s are the $SU(N_c)$ generators 
with $\tr{ \tau^a \tau^b } = 2 \delta^{ab}$.
In the Yukawa interaction,
$(n, \bar n)$ represent
pairs of hot spots 
connected by $\vec Q_{AF}$ :
$\bar 1 = 3, \bar 2 = 4, \bar 3=1, \bar 4=2$.
$\mu$ is an energy scale introduced
for the Yukawa coupling and the quartic couplings
which have the scaling dimensions $(3-d)/2$ and $(3-d)$ respectively.
For $N_c \leq 3$, $\tr{ \Phi^4} = \half \lt( \tr{ \Phi^2}
\rt)^2$ and $u_2$ is not an independent coupling.
In this case, it is convenient to set $u_2=0$ without loss of generality.
For $N_c \geq 4$, however, $u_1$ and $u_2$ are independent,
and one should keep both of them.
It is straightforward to check that 
\eq{eq:3D_theory} is reproduced from 
\eq{eq: generalD_theory} once we set $d=2$, $N_c=2$
and $N_f=1$.

\begin{figure}[!t]
\centering
 \includegraphics[width=0.7\textwidth]{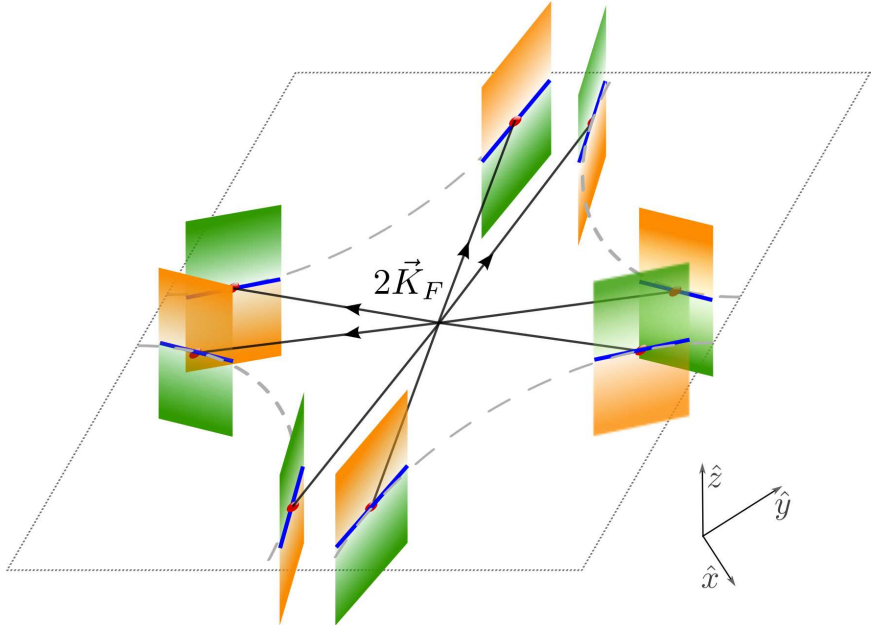}
 \caption{
One-dimensional Fermi surfaces embedded
in the three dimensional momentum space.
The locally flat patches of two-dimensional Fermi surface 
near the hot spots are gapped out by 
the $p_z$-wave charge density wave carrying momentum $2 \vec k_F$
except for the line nodes at $k_z$=0.
}
\label{fig:pCDW}
\end{figure}

\begin{figure}[!t]
\centering
 \includegraphics[width=0.5\textwidth]{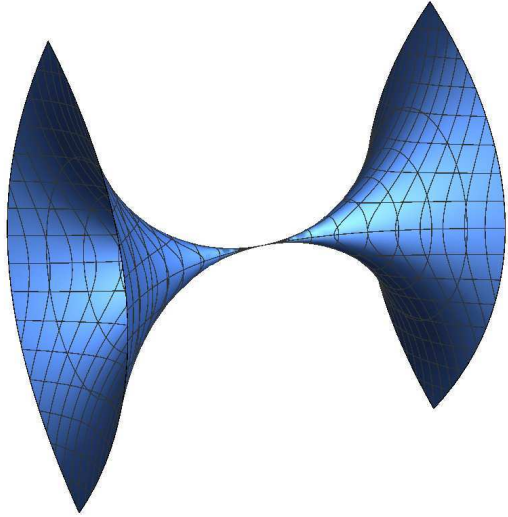}
 \caption{
A patch of Fermi surface created by a $p_z$-wave CDW with momentum $2 \vec k_F$.
The center of the patch is pinched due to the CDW order that vanishes linearly in $p_z$.
If the antiferromagnetic order parameter connects the pinched points,
the low energy effective theory for the critical point becomes
\eq{eq: generalD_theory} in three dimensions.
}
\label{fig: curvedFS}
\end{figure}

The action supports 
one-dimensional Fermi surfaces embedded
in $d$-dimensional momentum space.
The fermions have energy,
$E_n(k_1,..,k_{d-2},\vec k) = \pm \sqrt{ \sum_{i=1}^{d-2} 
k_i^2 + [\varepsilon_n(\vec k)]^2   }$
which disperses linearly in the $(d-1)$-dimensional space
perpendicular to the line node defined by 
$k_{i}=0$ for $1 \leq i \leq d-2$ and
 $\varepsilon_n(\vec k)=0$.
To understand the physical content of the dimensional regularization,
it is useful to consider the theory at $d=3$.
With the choice of 
$(\gamma_0, \gamma_1, \gamma_2) = ( \sigma_y, \sigma_z, \sigma_x)$
and identifying $k_1 = k_z$,
the kinetic energy 
for $\Psi_{1,\sigma}$ and $\Psi_{3,\sigma}$ 
is written as
$ H_0 =  
( v k_x \pm k_y )
\left[ 
  \psi^{(\pm)*}_{1,\sigma,j} \psi^{(\pm)}_{1,\sigma,j} 
- \psi^{(\pm)*}_{3,\sigma,j} \psi^{(\pm)}_{3,\sigma,j} 
\right]
\mp k_z \left[ \psi^{(\pm)*}_{1,\sigma,j}
\psi^{(\pm)}_{3,\sigma,j} + h.c. \right]$.
The kinetic energy
for $\Psi_{2,\sigma}$ and $\Psi_{4,\sigma}$ 
can be obtained by $90^\circ$ rotation.
The first term gives patches of locally flat two-dimensional Fermi surface.
The second term describes a $p_z$-wave charge density wave (CDW) that gaps out the two-dimensional Fermi surface
to leave line nodes at $k_z=0$, as is shown in Fig. \ref{fig:pCDW}. 
The full action in \eq{eq: generalD_theory}
describes the AF transition
driven by electrons near the hot spots
on the line nodes.

We emphasize that 
\eq{eq: generalD_theory} 
is not just a mathematical construction.
The theory in three space dimensions 
can arise at the AF quantum critical point
in the presence of $p_z$-wave CDW of momentum $2 \vec k_F$. 
If local curvature of the underlying Fermi surface is included, 
the dispersion near a pair of points on the Fermi surface connected by the momentum $2 \vec k_F$ 
can be written as
$\eps_{\pm}(\vec k) = \pm k_x + \gamma_1 k_y^2 + \gamma_2 k_z^2$, 
where $k_x$ is chosen to be perpendicular to the Fermi surface,
and $\gamma_i$ represent the local curvatures of the Fermi surface.
The $p_z$-wave CDW leads to the spectrum, $\mc{E}$
which is determined by 
$\begin{vmatrix}
 k_x + \gamma_1 k_y^2 + \gamma_2 k_z^2 - \mc{E} & k_z \\
k_z &  - k_x + \gamma_1 k_y^2 + \gamma_2 k_z^2 - \mc{E}
\end{vmatrix}
= 0. $
This results in a pinched Fermi surface 
located at $\gamma_1 k_y^2 + \gamma_2 k_z^2 = \sqrt{k_x^2 + k_z^2}$ as is shown in \fig{fig: curvedFS}. 
If the antiferromagnetic ordering wave vector connects
the pinched points, 
the low energy effective theory
for the phase transition is precisely described by 
\eq{eq: generalD_theory} in three dimensions.
Because the curvature is irrelevant at low energies,
the pinched Fermi surfaces can be regarded as Fermi lines
near the hot spots.
Similar field theory can also arise at an
orbital selective antiferromagnetic 
quantum critical point in three-dimensional 
semi-metal as is discussed in Appendix \ref{app: lattice}. 
The action in general dimensions respects 
the $U(1)$ charge conservation,
the $SU(N_c)$ spin rotation,
the $SU(N_f)$ flavor rotation,
the $90^\circ$ space rotation in $(k_x,k_y)$,
the reflections,
and the time-reversal symmetries.
For $N_c=2$, the pseudospin symmetry,
which rotates $\Psi_{n,\sig,j}(k)$ 
into $i \tau^{(y)}_{\sig, \sig'}~ \gamma_0  \bar{\Psi}^{T}_{n,\sig',j}(-k) $,
is present\cite{PhysRevB.82.075128}.
The action in \eq{eq: generalD_theory} is also invariant 
under the $SO(d-1)$ rotation in ${\bf K}$. 
In Appendix \ref{app: symm} we provide further details on symmetry.

The  theory in $2 \leq d \leq 3$ 
continuously interpolates the physical theories 
which describe the AF critical points in $d=2$ and $3$.
Because the couplings are marginal in three dimensions, 
we consider $d=3-\epsilon$ 
and expand around three space dimension 
using $\epsilon$ as a small parameter.
We use the field theoretic renormalization group scheme  
to compute the beta functions which
govern the RG flow of the renormalized 
velocities and coupling constants.


By embedding the one-dimensional Fermi surface in  higher dimensions,
the density of state (DOS) is reduced to $\rho(E) \sim E^{d-2}$.
As is the case for the usual dimensional regularization scheme for relativistic field theories,
the reduced DOS tames quantum fluctuations at low energies
and allows us to access low energy physics in a controlled way.
Of course, there is no guarantee that the physics obtained near $d=3$ is
continuously extrapolated all the way to $d=2$ because of the possibility
that some operators that are irrelevant near $d=3$ become 
relevant to drive instability near $d=2$.
However, it is our very goal to systematically 
examine the potential instability 
as dimension is lowered toward $d=2$,
for which we first need to establish the existence of stable fixed point
at $d=3$, which can be realized on its own.

{\bf Strange metal fixed point}.
We include one-loop quantum corrections 
to obtain the beta functions for the velocities and couplings 
(see Appendices \ref{app: RG} and \ref{app: 1Loop} for computational details), 
\begin{align}
\frac{dv}{dl} &=    - \frac{(N_c^2-1)}{2 N_c N_f \pi^2}~ 
\frac{z ~ v ~  g^2}{c} ~ h_2(v,c), 
\label{eq: beta_v_1L} \\
\frac{dc}{dl} &=   - \frac{z g^2}{8 \pi^2}~ 
\lt[ \pi  \frac{c}{v} 
- \frac{2(N_c^2-1)}{N_c N_f} [ h_1(v,c) - h_2(v,c) ] \rt] , 
\label{eq: beta_c_1L} \\
\frac{dg}{dl} &=    \frac{z}{2} ~ g \lt[ \eps  -
\frac{g^2 }{4\pi v} 
- \frac{g^2}{4 \pi^3 N_c N_f c} \lt\{ 
2(N_c^2-1) \pi ~h_2(v,c) 
- h_3(v,c) \rt\} \rt],  
\label{eq: beta_g_1L} \\
\frac{du_1}{dl} &=    
\frac{z u_1}{2 c^2 \pi^2 }
\bigg[
  c^2 \pi  \lt( 
2 \pi \epsilon -\frac{g^2}{v} 
\rt) 
+ \frac{ (N_c^2-1)}{N_f N_c} c g^2 \{ h_1(v, c) - h_2(v, c) \} \nn
& -(N_c^2+7) u_1 
- \frac{2(2 N_c^2-3)}{N_c}  u_2 
- 3 \lt(1 + \frac{3}{N_c^2} \rt) \frac{u_2^2}{u_1}   
\Bigg], 
\label{eq: beta_u1_1L}     \\
\frac{du_2}{dl} &=    
\frac{z u_2}{ 2 c^2 \pi^2  }
\Bigg[
  c^2 \pi \lt( 
 2 \pi  \epsilon 
-\frac{g^2}{v} 
\rt)
 + \frac{(N_c^2-1)}{N_f N_c} c g^2  \{ h_1(v, c) - h_2(v, c) \}
 -12 u_1 
- \frac{2(N_c^2 -9)}{ N_c} u_2  
\Bigg].
\label{eq: beta_u2_1L}    
\end{align}
Here
$l$ is the logarithmic length scale.
$z = \lt[ 1 - \frac{(N_c^2-1)}{4 N_c N_f \pi^2}~ 
\frac{g^2}{c}~ 
\Bigl\{ h_1(v,c)  - h_2(v,c) \Bigr\} \rt]^{-1}$ is the dynamical critical exponent
that determines the scaling dimension of ${\bf K}$ 
relative to $\vec k$. 
$h_i(v,c)$ are given by
$h_1(v, c) =  \int_0^1 dx ~ \sqrt{\frac{1-x}{c^2 + (1 + v^2
- c^2) x}}$,
$h_2(v, c) = c^2 \int_0^1 dx ~ 
\sqrt{\frac{1-x}{\lt[ c^2 + (1 + v^2 - c^2) x \rt]^3}}$
and $ h_3(v,c) =  \int_0^1 dx_1 \int_0^{1-x_1} 
dx_2 \int_0^{2\pi} d\theta \lt[ \frac{1}{\zeta(v, c, x_1, 
x_2, \theta)} -  \frac{v}{c} \frac{ \sin{2 
\theta}}{\zeta^2(v,c, x_1, x_2, \theta)}\rt]$
with
$$\zeta(v,c, x_1, x_2, \theta) 
 =  \frac{2v}{c} ( x_1 \cos^2{\theta} + x_2 \sin^2{\theta} ) 
+ (1 - x_1 - x_2)\lt[ v c \cos^2(\theta + 
\pi/4) + \frac{c}{v}  \sin^2(\theta + \pi/4)\rt].$$ 

\begin{figure}[!t]
\centering
\begin{subfigure}{0.65\textwidth}
\includegraphics[width=\columnwidth]{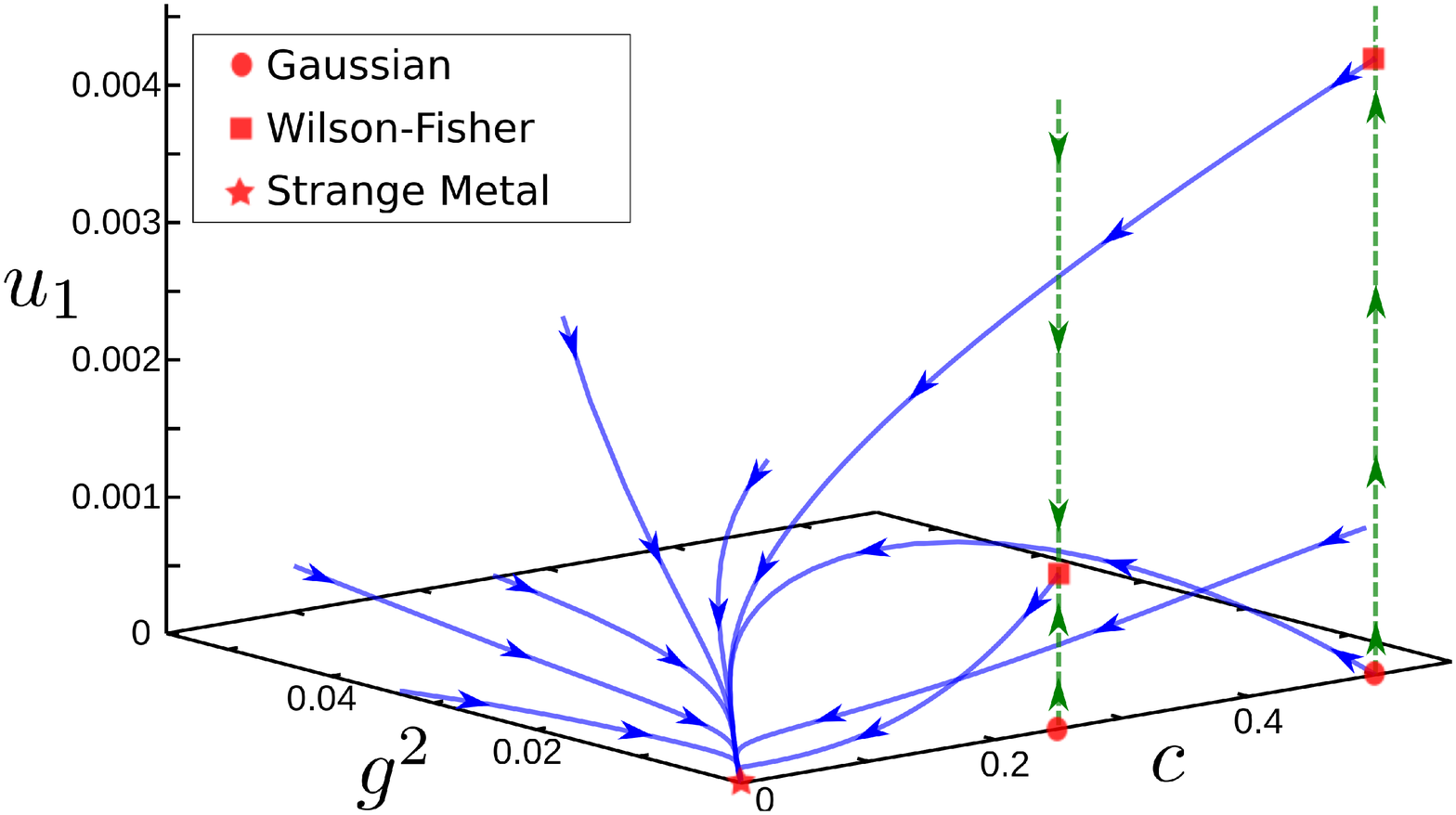}
\caption{}
\label{fig:flow}     
\end{subfigure}
\begin{subfigure}{0.65\textwidth}
\includegraphics[width=\columnwidth]{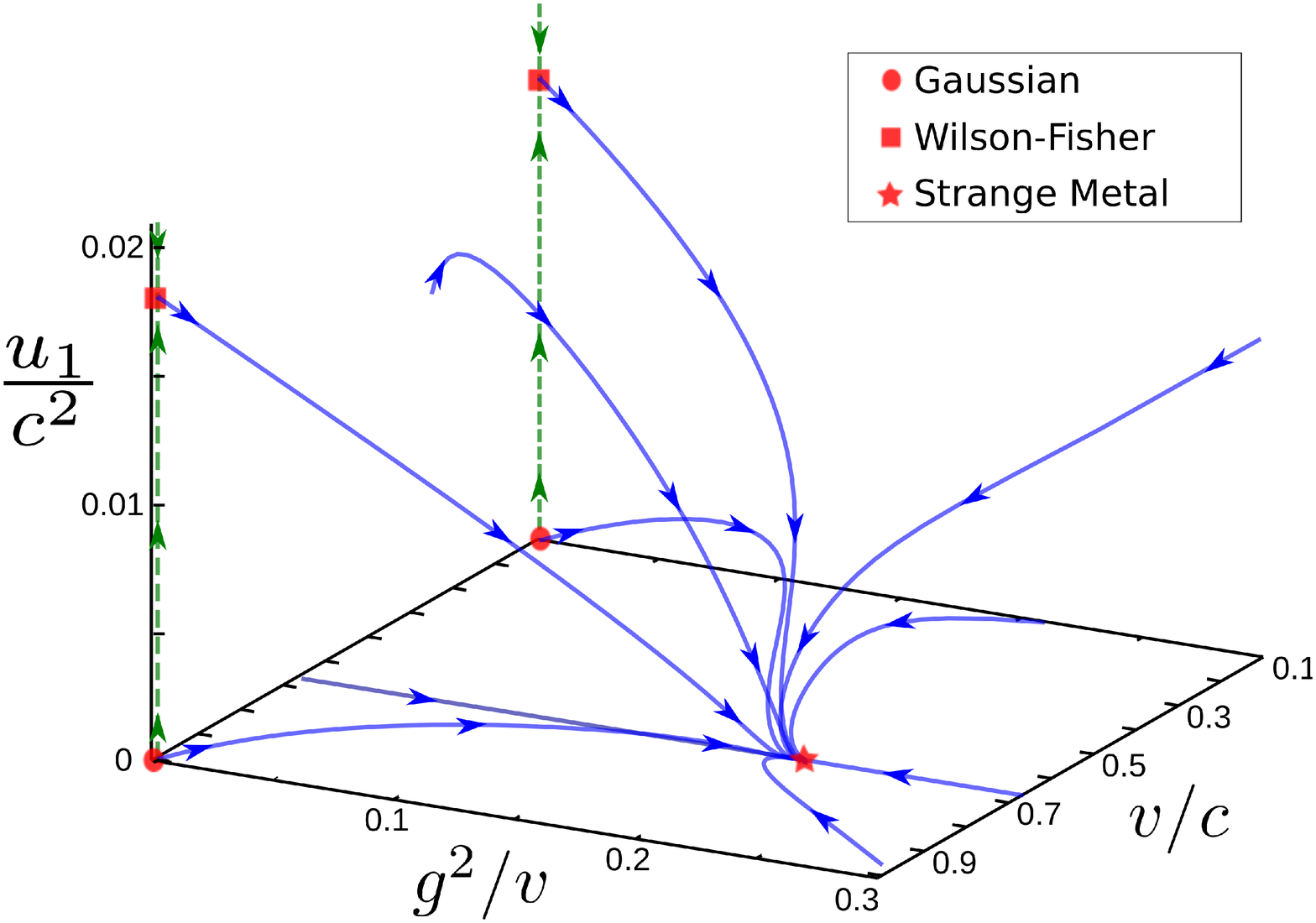}
\caption{}
\label{fig:ratio_flow}     
\end{subfigure}
 \caption{
One-loop RG flow of the couplings and velocities for $N_c=2$, $N_f=1$ and $\epsilon=0.01$.
We set $u_2=0$, which can be done without loss of generality for $N_c < 4$.
The solid lines denote flows in the three-dimensional space of the parameters shown in the figure.
The dashed lines represent flows within the subspace of $g=0$.
(a) The whole manifold of $g=u_1=0$ 
represents the non-interacting (Gaussian) fixed points
parameterized by $(v,c)$.
Once $u_1$ is turned on
at the Gaussian fixed points (denoted by circles),
the theory flows to the Wilson-Fisher fixed points (squares).
As the Yukawa coupling is introduced,
couplings and velocities flow to the stable fixed point at $g^*=u_i^*=v^*=c^*=0$.
(b) RG flow of the ratios of the parameters for the same
values of $N_c$, $N_f$ and $\epsilon$ as in (a).
The Yukawa coupling measured in the unit of $\sqrt{v}$ and the ratio
of the two velocities remain non-zero at the stable fixed
point.
}
\end{figure}

The RG flow of the couplings and the velocities
is shown in Fig. \ref{fig:flow}.
We first examine the RG flow in the subspace of $g=0$.
At the Gaussian fixed points ($u_i=g=0$ with $v,c \neq 0$), 
the theory is free. 
With $u_1 \neq 0$ and $u_2=0$, 
the theory flows to the $O(N_c^2-1)$ Wilson-Fisher (WF) fixed points 
at $u_1^*=\frac{2 \pi^2 c^2 \epsilon}{N_c^2+7}$
with dynamical critical exponent $z=1$.
For $N_c \geq 4$, one also needs to consider $u_2$.
As $u_2$ is turned on,
the $O(N_c^2-1)$ WF fixed points 
become unstable and it shows a run-away flow\cite{PhysRevB.88.125116}
which suggests a first-order phase transition.

In the presence of Yukawa coupling,
a stable low energy fixed point arises.
If some components of the velocities were not allowed to flow,
the theory could flow to a fixed point with finite couplings\cite{PhysRevB.87.045104}.
However, the full RG flow is more complicated because of running velocities.
As $g$ is turned on,
it initially grows 
as is expected from the fact that
it is relevant below $d=3$.
As $g$ grows,
the fermions at different hot spots 
are mixed with each other through quantum fluctuations.
As a result, the hot spots become increasingly nested at low energies : 
$v$ flows to zero as $1/l$ for $\epsilon >0$
and as $1/log(l)$ for $\epsilon =0$
in the low energy limit. 
The dynamical nesting of the fermionic band,
in turn, modifies the AF mode 
in two important ways.
First, the boson becomes increasingly slow
in the $q_x, q_y$ directions because the collective mode
can decay into dispersionless
particle-hole pairs near the nested hot spots.
As a result, $c$ decreases toward zero,
leading to emergent locality in the $(x,y)$ space.
Second, quantum fluctuations
become more and  more efficient
in screening the interactions 
due to the abundant low energy
density of states supported by 
the nested Fermi surface
and the dispersionless boson.
In summary of the RG flow,
i) $g$ induces dynamical nesting, renormalizing $v,c$ to smaller values,
ii) smaller $v,c$ make screening more efficient, making $g,u_i$ smaller.
This cycle of negative feedback leaves no room 
for a coexistence of the kinetic terms $(v,c)$ 
and the interactions $(g,u_i)$.
It has only one fate down the road of RG flow : mutual destruction.
To the one-loop order, 
all of $v, c, g, u_i$ eventually flow to zero 
in the low energy limit
at and below three dimensions
if initial values of $u_i$'s are not too large in magnitude. 

The new interactionless and quasi-dispersionless fixed point 
is distinct from the Gaussian fixed point 
which is dominated by the kinetic energy.
Unlike at the Gaussian fixed point,
the kinetic energy and the interactions maintain `a balance
of power'
along the path to their demise.
This can be seen from the fact that
the ratios defined by
\bqa
w \equiv \frac{v}{c}, 
~~
\lambda \equiv \frac{g^2}{v},
~~
\kappa_i \equiv \frac{u_i}{c^2}
\label{eq:ratios}
\eqa
flow to a stable fixed point, 
\bqa
w^* = \frac{N_c N_f}{N_c^2-1}, 
~~
\lambda^* = \frac{ 4 \pi (N_c^2+N_c N_f -1) }{ N_c^2+N_c N_f -3 }  \epsilon,
~~
\kappa_i^* = 0
\label{eq:fp}
\eqa
in the $c \rightarrow 0$ limit as is shown in Fig. \ref{fig:ratio_flow}.
At the fixed point, 
the dynamical critical exponent is renormalized to
$z = 1+\frac{\lambda^*}{8 \pi}$ to the leading order in $\lambda$.
The non-trivial quantum correction to $z$
implies that the effect of interaction
is not gone even though $g,u_i$ vanish in the low energy limit.
This is due to the emergent locality 
associated with the dynamical nesting of the Fermi surface
and the dispersionless bosonic spectrum.
The IR singularity supported by the locality 
makes the system infinitely susceptible to interaction,
leading to finite quantum corrections 
even with vanishing interactions.
The fixed point is stable for general $N_c$ and $N_f$,
and small perturbations of $w$, $\lambda$, $\kappa_i$ away from 
\eq{eq:fp} die out in the low energy limit.
In particular, the $\phi^4$ vertices acquire an anomalous dimension
and become irrelevant at the new fixed point.
The one-loop fixed point is exact at $d=3$
because higher order terms are systematically
suppressed by $\lambda$ and $\kappa_i$ 
which flow to zero in the low energy limit.
For $d < 3$, $\kappa_i$, $c$, $v$ 
can receive higher-loop corrections 
to become nonzero at the fixed point.
The details on higher-loop contributions
can be found in Appendix \ref{app: higherLoops}.

If the initial value of $\kappa_1$ is sufficiently large and negative, 
$\kappa_1$ runs away to $-\infty$, potentially driving a first-order transition.
The stable fixed point in \eq{eq:fp}  
and the run-away flow is separated
by an unstable fixed point at
$\kappa_1^* = - \frac{ 4\pi^2 \epsilon}{(N_c^2+7)(N_c^2 + N_c N_f - 3)}, \kappa_2^*=0$
with the same values of $w^*$ and $\lambda^*$ 
as in \eq{eq:fp}. 
The unstable fixed point, which can be realized at a multi-critical point,
describes a state  distinct from the state described by the stable 
fixed point  in \eq{eq:fp}. 
The two fixed points are distinguished by
the different ways the couplings and velocities approach the origin.

{\bf Physical properties}.
The existence of the stable low energy fixed point implies
scale invariance of the Green's function 
in the limit 
$k_x, k_y, |{\bf K}|$ go to zero 
with $\frac{k_y}{k_x}$,  $\frac{{\bf K}}{ |k_x|^z }$ fixed
at the second order phase transition.
Here we focus on the Green\rq{}s function near the hot spot $1+$
in Fig. \ref{fig:hot_spots}.
The Green\rq{}s function near other hot spots can be obtained 
by applying reflection or $90^\circ$ rotation. 
In the scaling limit, 
the fermion Green\rq{}s function takes the form,
\bqa
{\cal G}(k) & = &  \frac{1}{ | k_y |^{1-2 \td \eta_\psi} } 
\td G \left(  \frac{ {\bf  K} }{ |k_y|^z }    \right), 
\label{eq:G}
\eqa
where $\td \eta_\psi \sim O(\epsilon^2)$ is the net anomalous dimension 
which vanishes to the linear order in $\epsilon$
and $\td G(x)$ is a universal function.
Because $v$ flows to zero logarithmically in the low energy limit,
the dependence on $k_x$ is suppressed as $\frac{k_x}{log(1/k_x)}$ for $d<3$
and as $\frac{k_x}{log\lt( log(1/k_x)\rt)}$ at $d=3$ in the scaling limit.
The dynamical critical exponent is non-trivial even to the linear order in $\epsilon$ for $d<3$.
As a result, the spectral function 
shows a power-law distribution in energy 
instead of a delta function peak,
exhibiting a non-Fermi liquid behavior.
%
At $d=3$, we have $z=1$ as in Fermi liquid.
However, the Green\rq{}s function is modified by logarithmic corrections
compared to that of the Fermi liquid
due to $\lambda$ which flows to zero logarithmically.
This is a marginal Fermi liquid\cite{varma1989phenomenology}.
Since the boson velocity also flows to zero in the same fashion $v$ flows to zero, 
the boson Green's function becomes independent of $k_x, k_y$ in the scaling limit
upto corrections that are logarithmically suppressed,
\bqa
{\cal D}(k) & = &  \frac{C}{ | {\bf  K} |^{\frac{2-2 \td \eta_\phi}{z} } },
\label{eq:D}
\eqa
where $C$ is a constant and $\td \eta_\phi  \sim O(\epsilon^2)$.
This {\it quasi-local strange metal} 
supports non-quasiparticle excitations 
which are dispersionless along $x, y$ directions 
in the scaling limit.
Here the effective space dimension
becomes dynamically reduced as a result of quantum fluctuations.
The quasi-local behaviors associated with extreme velocity anisotropies 
were reported in nodal semi-metals\cite{PhysRevB.78.064512,2014arXiv1403.5255S}.
Local critical behaviors with $z=\infty$
also arise in the dynamical mean-field approximation
for the Kondo lattice model\cite{si2001locally}
and from gravitational constructions\cite{PhysRevD.79.086006,vcubrovic2009string,PhysRevD.83.125002}.
The present quasi-local state 
is distinct from the earlier examples
in that it is a stable zero temperature state
which supports extended Fermi surface with a finite $z$.


The quasi-local strange metal is stable 
at the one-loop order
which becomes exact in the $d \rightarrow 3$ limit. 
As one approaches $d=2$, higher order corrections become important.
The theory at $d=2$ remains strongly coupled 
even in the large $N_c$ and/or large $N_f$ limit.
One possibility is that the quasi-local strange metal
becomes unstable towards an ordered state below a critical dimension.
To identify the channels that may become unstable at $d=2$,
we examine charge density wave (CDW) and superconducting (SC)
correlations that are enhanced by quantum fluctuations\cite{PhysRevB.82.075128,Berg21122012}. 
In principle, particle-hole or particle-particle fluctuations
between un-nested patches of Fermi surface may drive an instability
if the coupling is strong at the lattice scale\cite{efetov2013pseudogap}.
However, those operators that connect nested patches
receive strongest quantum corrections.

In the spin-singlet CDW channel, the set of  operators
$$O_{CDW}^{\pm}  =  \int dk \lt[ 
\lt(
  \bar \Psi_{1,\sigma,j} \Psi_{1,\sigma,j}
+ \bar \Psi_{3,\sigma,j} \Psi_{3,\sigma,j}
\rt) 
\pm
\lt(
  \bar \Psi_{2,\sigma,j} \Psi_{2,\sigma,j}
+ \bar \Psi_{4,\sigma,j} \Psi_{4,\sigma,j}
\rt) \rt] $$
which describes a $p_y$-wave and a $p_x$-wave CDW, respectively, with momentum $2 \vec k_F$
is most strongly enhanced.
These CDW operators, which are pseudospin singlets for $N_c = 2$, break the reflection symmetry and 
represent bond density waves without on-site modulation of charge. This is different from the bond density wave order which forms a pseudospin doublet with the d-wave pairing order \cite{PhysRevLett.111.027202}.
In the SC channel, we focus on the representation
that is symmetric in $SU(N_f)$ and anti-symmetric in $SU(N_c)$,
which reduces to the spin-singlet SC order for $N_c=2, N_f=1$.
There are two sets of equally strong SC fluctuations.
The first set of operators describes 
the $d_{x^2-y^2}$-wave and $g$-wave pairings with zero momentum\cite{scalapino1986d, PhysRevB.34.6554},
while the second set of operators
describes $s$-wave and $d_{xy}$-wave pairings 
with finite momentum, $2 \vec
k_F$\cite{PhysRev.135.A550,LL}.
The attractive interaction for the pairing is 
mediated by the commensurate spin fluctuations
that scatter a pair of electrons from one hot spot
to another hot spot.
Due to the nesting, the finite momentum pairing is 
as strong as the conventional 
zero momentum pairing
to the one-loop order.
The propensity for finite momentum pairing may 
lead to exotic superconducting states
in two dimensions\cite{berg2009charge,2014arXiv1401.0519L}.
If the quasi-local strange metal is unstable 
toward a competing order at low temperature in two dimensions,
the strange metallic behaviors predicted in Eqs. (\ref{eq:G}) and (\ref{eq:D})
can show up within a finite temperature window
whose range can be made parametrically large 
by tuning $N_c$ and $N_f$\cite{PhysRevB.89.165114}.
For more details on the computation of 
the anomalous dimensions for the CDW and SC orders, 
please see Appendix \ref{app: suscep}.

Within the perturbative regime that we explore in this paper, 
the anomalous dimensions for various susceptibilities associated with `hot' electrons near the hot spots remain small.
Because hot spots are only points in momentum space,
thermodynamic and transport properties are dominated by cold electrons which exhibit Fermi liquid behaviors. 
For example, the specific heat will be proportional to $T^{2-\epsilon}$ to the leading order of temperature $T$, 
and the conductivity is expected to be dominated by cold electrons\cite{hartnoll2011quantum}.
As one approaches $d=2$, 
the contribution from hot electrons may, in principle, dominate over 
the contribution from the cold electrons as the anomalous dimensions become larger. 
Moreover, the behavior of cold electrons may also deviate from those of Fermi liquid far away from three dimensions,  
as the coupling between cold electrons and collective modes, which is irrelevant in the perturbative regime, becomes 
strong near $d=2$\cite{hartnoll2011quantum}. 
In this case, non-Fermi liquid behavior may show up even for the thermodynamic and transport properties of cold electrons. 
However, we can not address this issue in a controlled manner because it requires strong coupling which lies outside the perturbative window.

{\bf Conclusion}.
We show that a novel strange metallic state
emerges at the AF quantum critical point 
in a metal that supports one-dimensional Fermi surface
based on a perturbative expansion
which gives the exact low energy fixed point 
in three dimensions. 
Even though the interaction 
is screened to zero in the low energy limit,
dynamical reduction of the effective dimensionality
drives the system into a strange metallic state,
which supports partially dispersionless incoherent single-particle excitations
along with enhanced superconducting and charge density wave fluctuations.
The present theory continuously interpolates between 
the three dimensional theory for one dimensional Fermi surfaces
and two dimensional metals.
The three-dimensional theory can arise at the AF quantum critical point
in the presence of $p_z$-wave CDW, 
which is described by a stable quasi-local marginal Fermi liquid.
Our formalism also provides a way to access potential instabilities of the 
non-Fermi liquids that arise at the AF quantum critical points 
below three dimensions as $\epsilon$ increases.

\begin{acknowledgments}
We thank Andrey Chubukov, Catherine Pepin, Patrick Lee, Max Metlitski, Subir Sachdev, {T . Senthil} and Yong Baek Kim for helpful discussions.
 The research was supported in part by 
 the Natural Sciences and Engineering Research Council of 
 Canada, the Early Research Award from the Ontario Ministry of Research and Innovation, and the Templeton Foundation.
 Research at the Perimeter Institute is supported 
 in part by the Government of Canada 
 through Industry Canada, and by the Province of Ontario through the Ministry of Research and Information.
\end{acknowledgments}


\appendix

%
%
%
%
%
%
%
%
%
%

\section{A three-dimensional lattice model for a related field theory} \label{app: lattice}
\begin{figure}[!ht]
\centering
\begin{subfigure}[b]{0.47\textwidth}
\includegraphics[width=0.8\columnwidth]{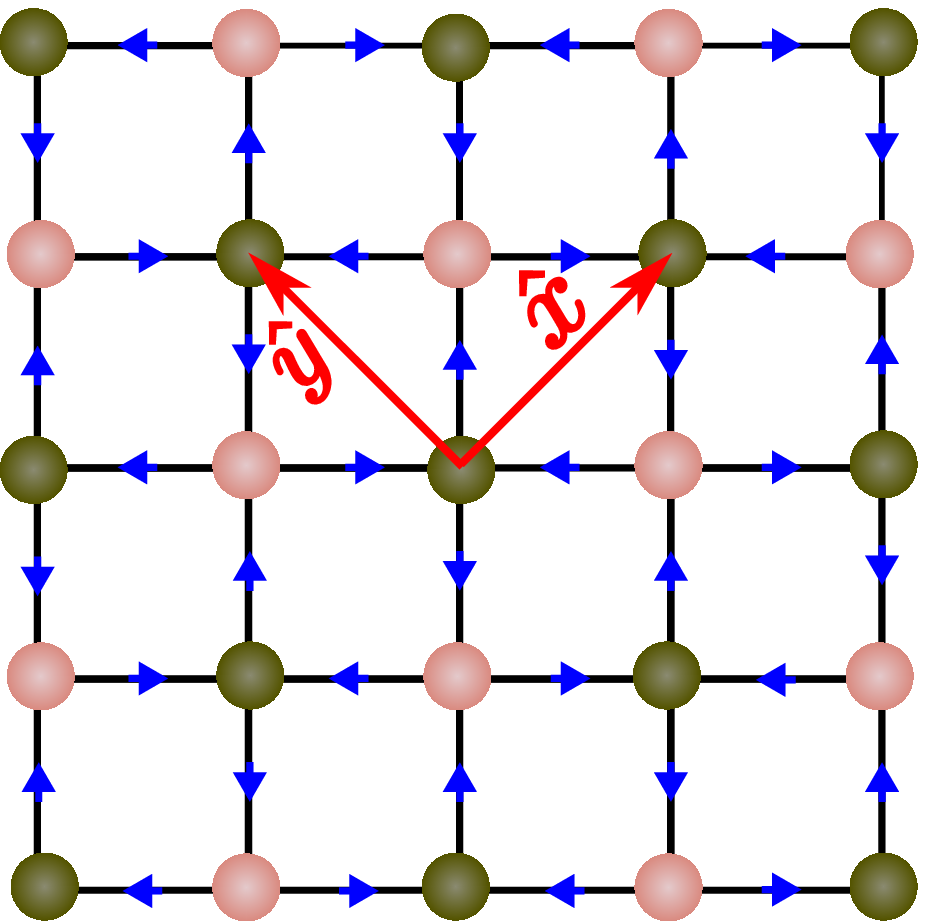}
\caption{}
\label{fig: flux_001}
\end{subfigure}
\begin{subfigure}[b]{0.47\textwidth}
\includegraphics[width=0.9\columnwidth]{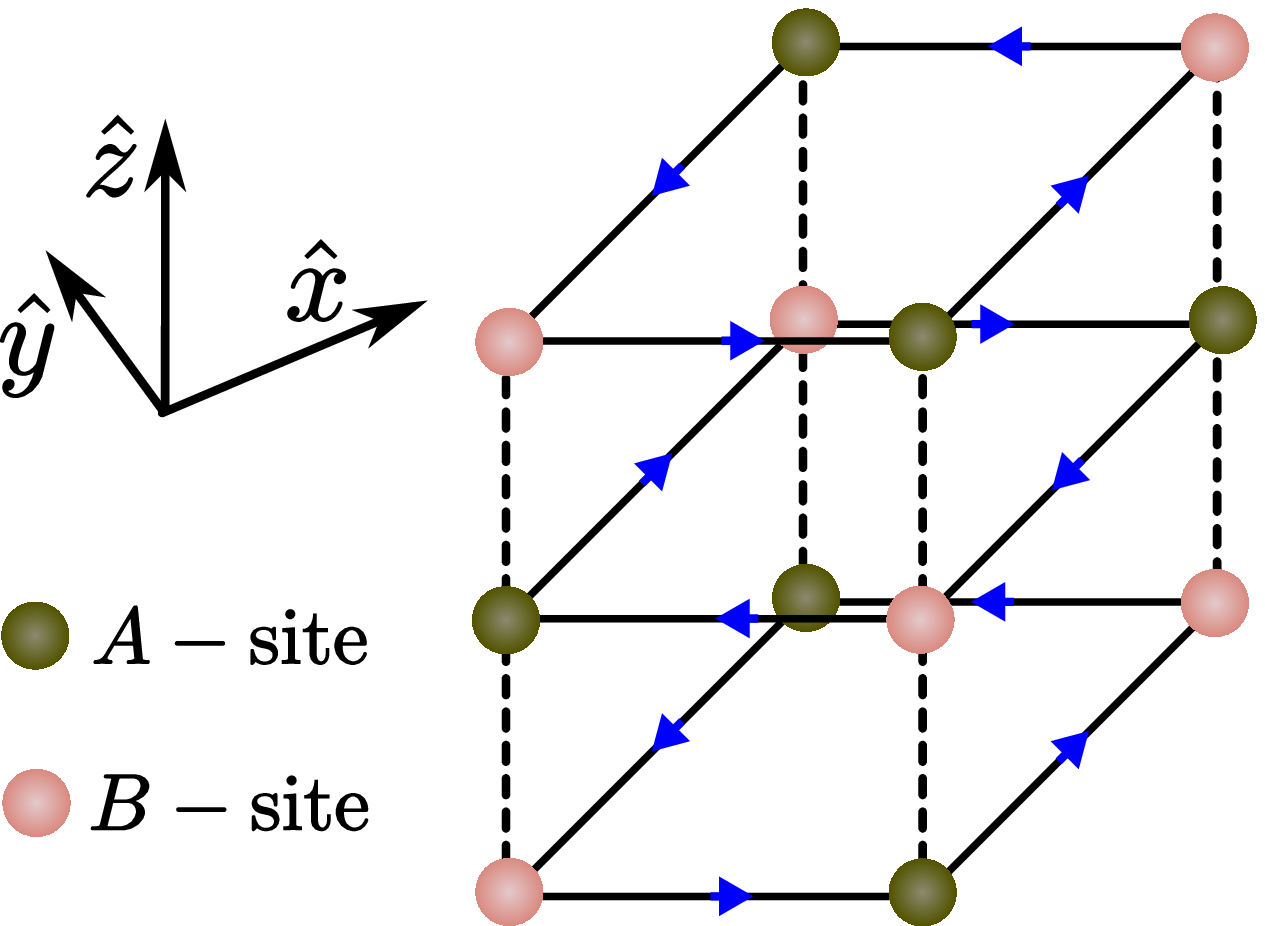}
\caption{}
\label{fig: flux_3D}
\end{subfigure}
\caption{
(a) Flux lattice in the $XY$-plane. 
The green (dark) and the red (light) disks represent sites $A$ and $B$, respectively. 
(b) Three dimensional arrangement of the $A$ and $B$ sites.}
\end{figure}

In this section we construct a three dimensional lattice model
in which a field theory similar to the one considered in the main text
can be realized.
We consider a tetragonal lattice
where staggered fluxes 
pierce through unit plaquettes.
A gauge is chosen such that
the hopping $t_z$ along the $z$ direction is real.
The nearest neighbor hoppings in the $XY$-plane are written as
$\{ t_{1}  e^{i \phi}, t_{2}  e^{i \phi} \}$ 
$( \{t_{1}  e^{-i \phi}, t_{2}  e^{- i \phi}\} )$ 
in the two orthogonal directions along (against) the arrows,
as is shown in Figs. \ref{fig: flux_001} and \ref{fig: flux_3D}.
Here the magnitudes of staggered flux per plaquette are $(4 \phi, 2\phi, 2\phi)$ 
in the three planes.
In the coordinate system shown in
Figs. \ref{fig: flux_001} and \ref{fig: flux_3D},
the lattice vectors become
$\vec a_1 = (1,0,0)$, 
$\vec a_2 = (0,1,0)$, 
$\vec a_3 = \frac{1}{2}(1,1,1)$,
where the distance between nearest neighbor $A$ sites in the $XY$-plane is chosen to be $1$.
The reciprocal vectors are given by 
$\vec b_1 = 2\pi (1,0,-1)$, 
$\vec b_2 = 2\pi (0,1,-1)$, 
$\vec b_3 = 4\pi (0,0,1)$.

\begin{figure}[!ht]
\centering
\includegraphics[width=0.6\columnwidth]{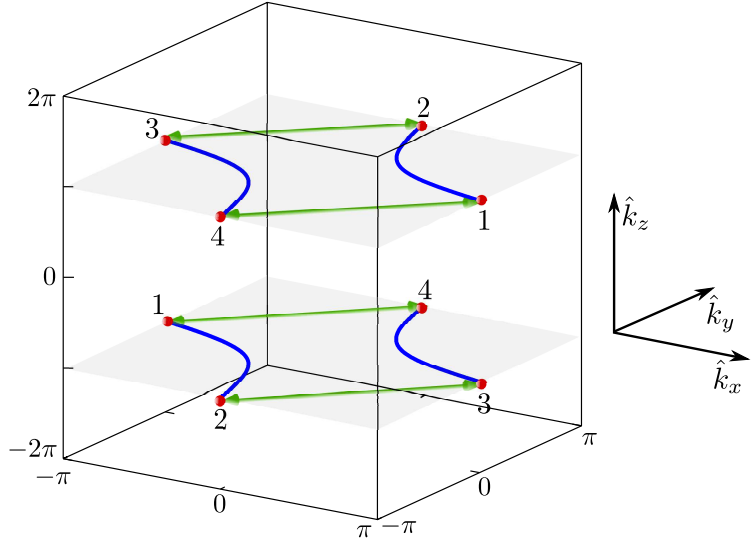}
\caption{ 
The (blue) curves represent the Fermi lines embedded in the three dimensional momentum space for \eq{eq: nodal_FS} with $t_{1} = 0.6 ~t_{2}$. 
The (green) arrows represent the antiferromagnetic ordering vector with $\vec Q = (\pi,\pi,0)$. 
The shaded regions represent the $k_z= \pm \pi$ planes. 
There are four distinct hot spots connected by $\vec Q$ denoted by $n=1, 2, 3, 4$.
}
\label{fig: FermiLine}
\end{figure}

The tight binding Hamiltonian with the nearest neighbor hoppings becomes
\begin{align}
H &= -\sum_{\vec k} 
\left[
\mc{D}(\vec k) ~ 
c^{\dag}_A(\vec k) ~ 
c_B(\vec k) + h.c. \right],
 \label{eq: full_Hamiltonian}
\end{align}
where $c_{A(B)}$ is the destruction operator for electrons at $A$ ($B$) sites, and 
\begin{align}
\mc{D}(\vec k) & = 2 \lt[ \cos(\phi) \lt\{ t_+  \cos\lt( \frac{k_x}{2} \rt) \cos\lt( \frac{k_y}{2} \rt) + t_-  \sin\lt( \frac{k_x}{2} \rt) \sin\lt( \frac{k_y}{2} \rt) \rt\}+  t_z \cos\lt( \frac{k_z}{2} \rt) \rt] \nn
& \qquad  + 2 i  \sin(\phi) \lt\{ t_+ \sin\lt( \frac{k_x}{2} \rt) \sin\lt( \frac{k_y}{2} \rt) + t_- \cos\lt( \frac{k_x}{2} \rt) \cos\lt( \frac{k_y}{2} \rt) \rt\},
\end{align}
with $t_\pm = t_{1} \pm t_{2}$. 
The Hamiltonian is diagonal in spin indices,
and we have suppressed the spin indices in the electron operators.
The $2 \times 2$ Hamiltonian supports a particle-hole symmetric band
with the dispersion $E(\vec k)  = \pm \abs{\mc{D}(\vec k)}$ at half filling.
With the choice of $0< t_{1} < t_{2}$ and $\phi = \frac{\pi}{2}$,
one obtains one-dimensional Fermi surfaces (or \textit{Fermi lines})  located at
\begin{align}
& k_z = \pm \pi,  \nn
& \sin\lt( \frac{k_x}{2} \rt) \sin\lt( \frac{k_y}{2} \rt) + \frac{t_-}{t_+} \cos\lt( \frac{k_x}{2} \rt) \cos\lt( \frac{k_y}{2} \rt)  = 0.
\label{eq: nodal_FS}
\end{align} 
The Fermi lines embedded in the three-dimensional momentum space 
are shown in \fig{fig: FermiLine}.

\begin{figure}[!ht]
\centering
\begin{subfigure}[b]{0.45\textwidth}
\includegraphics[width=0.8\columnwidth]{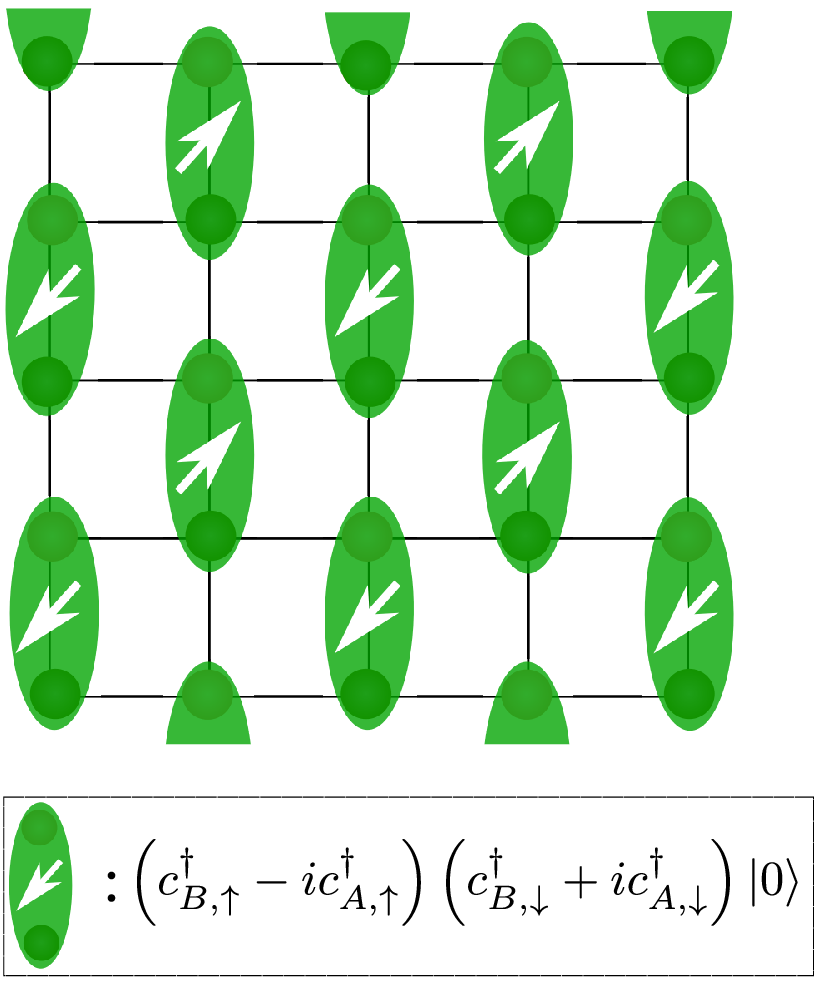}
\caption{}
\end{subfigure}
\begin{subfigure}[b]{0.49\textwidth}
\includegraphics[width=0.8\columnwidth]{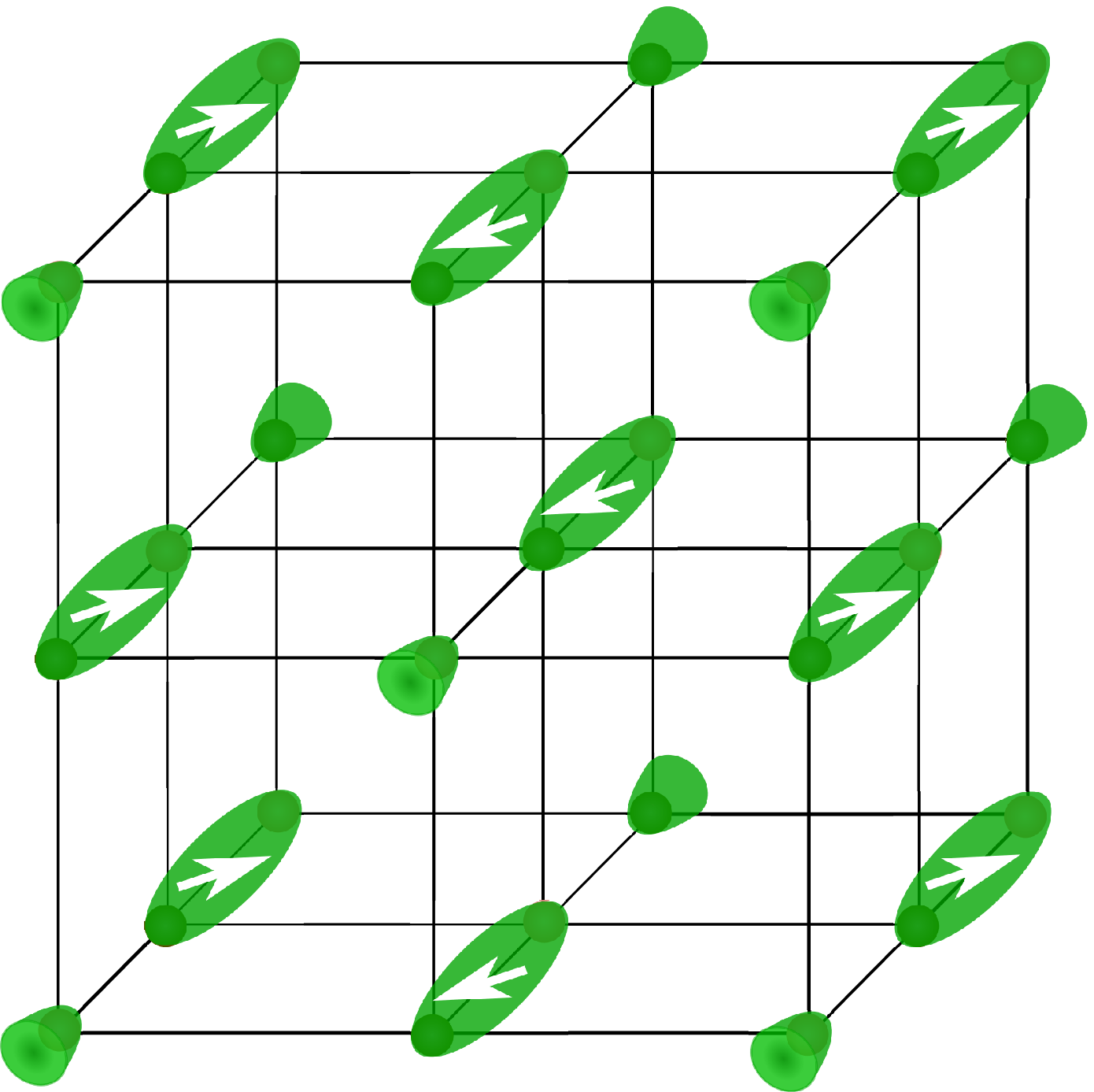}
\caption{}
\end{subfigure}
\caption{
A real space pattern for the orbital selective antiferromagnetic order (a) on the $XY$ plane and (b) in the full three dimensional lattice.
}
\label{fig: sdw}
\end{figure}

We assume that there exists an electron-electron interaction
which drives the semi-metal into an antiferromagnetic state.
In particular, we consider an orbital selective antiferromagnetic order, 
where electrons in the bonding and anti-bonding states within an unit cell 
have opposite spins, which then modulate 
with momentum $(\pi, \pi, 0)$ in space.
This is illustrated in Fig. \ref{fig: sdw}.
If the phase transition is continuous,
the critical spin fluctuations associated with the order
strongly interact with electrons on the Fermi lines
connected by the ordering vector $\vec Q = (\pi,\pi,0)$ 
as is shown in \fig{fig: FermiLine}. 
In this case, there exist four distinct hot spots 
connected by the ordering vector. As is considered in the main text, the minimal theory that describes the quantum critical point includes the electronic excitations near the hot spots and the critical antiferromagnetic mode 
that is coupled with electrons through
the Yukawa coupling,
\begin{align}
\mc{S}_{eff} &= \int \frac{d^4 k}{(2\pi)^4}  \sum_n \bar \Psi_n(k) \lt[ ik_0 \gamma_0 + i \gamma_1 k_z + i \gamma_2 \veps_n(\vec k) \rt] \Psi_n(\vec k) \nn 
& \quad + i g_0 \int \frac{d^4 k}{(2\pi)^4} \frac{d^4 q}{(2\pi)^4} ~ \vec \Phi(q) \cdot \lt[ \bar \Psi_{1;s}(k+q)  \gamma_2 \vec \tau_{s,s'} \Psi_{4;s'}(k) +  \bar \Psi_{2;s}(k+q) \gamma_2 \vec \tau_{s,s'} \Psi_{3;s'}(k) 
 \rt] + h.c. \nn 
& \quad  + \half \int \frac{d^4 q}{(2\pi)^4} \lt[q_0^2 + c_z^2 q_z^2 + c_x^2 q_x^2 + c_y^2 q_y^2 \rt] \vec \Phi(-q) \cdot \vec \Phi(q) \nn
& \quad + \frac{u_0}{4!} \int \frac{d^4 q_1}{(2\pi)^4}\frac{d^4 q_2}{(2\pi)^4}\frac{d^4 q_3}{(2\pi)^4} \lt[\vec \Phi(q_1 + q_2) \cdot \vec \Phi(q_2) \rt] \lt[ \vec \Phi(q_3 - q_2)  \cdot  \vec \Phi(q_3) \rt].
\label{eq: 3D_effective-2}
\end{align}
Here $\Psi_n(\vec k) \equiv 
\left( \begin{array}{c} 
c_A(\vec K_n + \vec k) \\
c_B(\vec K_n + \vec k) \end{array} \right)$
with $n=1,2,3,4$ denotes electrons near the $n$-th hot spot 
with
\begin{align}
& \vec K_1 = \lt(\pm \pi, 0, \pm \pi \rt), \quad 
\vec K_2 = \lt(0,\pm \pi, \pm \pi \rt), \quad
 \vec K_3 = \lt(\mp \pi, 0,\pm \pi \rt), \quad  
\vec K_4 = \lt(0, \mp \pi,\pm \pi \rt),
\end{align}
and
$\bar \Psi_n \equiv \Psi^{\dag}_n \gamma_0$ with  
$\gamma_0 = \sig_z$, $\gamma_1 = \sig_y$, and $\gamma_2 = \sig_x$. 
$\veps_n(\vec k)$ is the linearized dispersion around each hot spot,
\begin{align}
& \veps_1(\vec k) =  t_- k_x - t_+ k_y, \qquad 
\veps_2(\vec k) =  - t_+ k_x + t_- k_y, \nn
& \veps_3(\vec k) = - t_- k_x + t_+ k_y, \qquad  
\veps_4(\vec k) =  t_+ k_x - t_- k_y.
\end{align}
We have scaled away $t_z$ dependence by absorbing it in the $\hat z$-component of momentum. 
$\vec \Phi(q)$ represents the critical fluctuations of the SDW order parameter.

Unlike the action in \eq{eq: generalD_theory},
\eq{eq: 3D_effective-2} lacks 
the $C_4$ symmetry in the $XY$-plane and 
the $SO(2)$ rotation symmetry in the $k_0 - k_z$ plane.
This results in velocity anisotropy for the bosons.
However, we expect that this theory also flows to 
a quasi-local fixed point similar 
to the one discussed in the main text\cite{Sur2015}. 


\section{Symmetry} \label{app: symm}

In this section, we elaborate on the symmetries of the action 
in \eq{eq: generalD_theory}.
The internal symmetry is $U(1)^2 \times SU(N_c) \times SU(N_f)^2$  
associated with charge, spin and flavor conservations.
There are two $U(1)$'s and two $SU(N_f)$'s 
because the charge and flavor are conserved 
within the two sets of hot spots ($\{ 1, 3 \}$ and $\{2, 4 \}$) separately.
Besides the internal symmetry,
the action has $\pi/2$ rotation and reflection symmetries
under which the spinors transform as is shown in the Table \ref{tb:1}.
In $d>2$, the $SO(d-1)$ spacetime rotational symmetry is present.
Under $SO(d-1)$ rotation, $k_\mu$ and $\bar \psi_{n,\sigma,j} \gamma_\mu \psi_{n,\sigma,j}$
form vectors for $\mu=0,1,\ldots,d-2$.
For $N_c = 2$, there also exists a pseudo-spin symmetry under which
the super-spinor,
$ \chi_{n,\sig,j}(k) = 
\lt( \Psi_{n,\sig,j}(k), 
 i \tau^{(y)}_{\sig, \sig'}~ \gamma_0 ~ \bar{\Psi}^{T}_{n,\sig',j}(-k) \rt)^T
$
transforms as $\chi_{n,\sig,j} \mapsto U \chi_{n,\sig,j}$,
where $U$ represents $SU(2)$ matrix that acts on the particle-hole space.

\begin{table}[h]
\centering
\begin{tabular}{|c||c|c|c|}
\hline
      &	$R_{\pi/2}$  &	$R_x$	&	$R_y$ \\
\hline
\hline
$\Psi_{1}(k)$ & $\Psi_{2}(k_{R_{\pi/2}}) $ & $-i\gamma_0 \Psi_{3}(k_{R_x}) $ & $ i \gamma_0 \gamma_{d-1}  \Psi_{3}(k_{R_y})$  \\
\hline
$\Psi_{2}(k)$ & $\gamma_{d-1} \Psi_{1}(k_{R_{\pi/2}}) $ & $i\gamma_0 \gamma_{d-1} \Psi_{4}(k_{R_x})$ & $ -i \gamma_0 \Psi_{4}(k_{R_y})$  \\
\hline
$\Psi_{3}(k)$ & $\Psi_{4}(k_{R_{\pi/2}}) $ & $i\gamma_0 \Psi_{1}(k_{R_x})$ & $ i \gamma_0 \gamma_{d-1}  \Psi_{1}(k_{R_y})$  \\
\hline
$\Psi_{4}(k)$ & $-\gamma_{d-1} \Psi_{3}(k_{R_{\pi/2}}) $ & $i\gamma_0 \gamma_{d-1} \Psi_{2}(k_{R_x})$ & $ i \gamma_0  \Psi_{2}(k_{R_y})$  \\
\hline
\end{tabular}
\caption{
Table of spinors obtained by applying the spatial $\pi/2$ rotation 
and reflections in the $x$ and $y$ directions accompanied by
reflections in $k_1,k_2,\ldots,k_{d-2}$.
Under the three space symmetries,
the energy-momentum vector $ k=(k_0,k_1,\ldots,k_{d-2},k_x,k_y)$
is transformed to
$  k_{R_{\pi/2}}=(k_0,k_1,\ldots,k_{d-2},-k_y,k_x) $,
$  k_{R_{x}}=(k_0,-k_1,\ldots,-k_{d-2},-k_x,k_y) $ and
$  k_{R_{y}}=(k_0,-k_1,\ldots,-k_{d-2},k_x,-k_y) $, respectively.
The spin and flavor indices are suppressed.
}
\label{tb:1}
\end{table}

\section{Renormalization group analysis } \label{app: RG}

In this section, 
we describe the method that
is used to compute the beta functions 
for the velocities and couplings.
In order to incorporate quantum corrections,
we renormalize the theory 
by `tuning' the parameters in the action
in \eq{eq: generalD_theory}
such that the physical observables 
become insensitive to the UV cut-off scale.
This amounts to adding 
counter terms that 
remove UV divergences
in the quantum effective action 
order by order in the couplings.
The internal and spacetime symmetries guarantee that
the counter terms take the following form,
\begin{align}
  \mathcal{S}_{CT} &= 
\sum_{n=1}^4 
\sum_{\sigma=1}^{N_c} 
\sum_{j=1}^{N_f} 
\int \frac{d^{d+1}{k}}{(2\pi)^{d+1}} ~
\bar{\Psi}_{n,\sigma,j}(k) 
\lt[ i \mathcal{A}_1 \mathbf{\Gamma} \cdot \mathbf{K}  
+ i \mc{A}_{3} 
\gamma_{d-1} 
\veps_n\lt( \vec k; \frac{ \mc{A}_{2} }{ \mc{A}_{3} } v \rt)
\rt] 
\Psi_{n,\sigma,j}(k)
\nn 
&  + \frac{1}{4} 
 \int \frac{d^{d+1}q}{(2\pi)^{d+1}}
\Bigl[ \mathcal{A}_4 \abs{\mathbf{Q}}^2 + \mathcal{A}_5
~ c^2 \abs{\vec q}^2 \Bigr]
\tr{ \Phi(-q)  ~ \Phi(q) } \nn
  & + i \mc{A}_6 
\frac{g \mu^{(3-d)/2}}{\sqrt{N_f}}
\sum_{n=1}^4
\sum_{\sigma,\sigma'=1}^{N_c}
\sum_{j=1}^{N_f} 
\int \frac{d^{d+1}{k}}{(2\pi)^{d+1}} 
\frac{d^{d+1}{q}}{(2\pi)^{d+1}} 
\Bigl[
\bar{\Psi}_{\bar n,\sigma,j}(k+q) 
\Phi_{\sigma,\sigma'}(q)
\gamma_{d-1} 
\Psi_{n,\sigma',j} (k) 
\Bigr]  \nn
&  +  \frac{ \mu^{3-d} }{4}
\int 
\frac{d^{d+1}{k}_1}{(2\pi)^{d+1}} 
\frac{d^{d+1}{k}_2}{(2\pi)^{d+1}} 
\frac{d^{d+1}{q}}{(2\pi)^{d+1}}
 \Bigl[ 
\mc{A}_7 u_1 \tr { \Phi(k_1+q)  \Phi(k_2-q) } \tr{ \Phi(k_1)  \Phi(k_2) } \nn
& + \mc{A}_8 u_2 \tr { \Phi(k_1+q)  \Phi(k_2-q)  \Phi(k_1)  \Phi(k_2) } \Bigl], 
\label{eq: CT_action_general}
\end{align}
where  
$\veps_1(\vec k;v) = v k_x + k_y$, 
$\veps_2(\vec k;v) = - k_x + v k_y$, 
$\veps_3(\vec k;v) =  v k_x - k_y$, and
$\veps_4(\vec k;v) = k_x + v k_y$. 
In the minimal subtraction scheme, 
the counter terms only include contributions
that are divergent in the $\epsilon \rightarrow 0$ limit,
\begin{align}
\mathcal{A}_n \equiv \mathcal{A}_n(v,c,g,u;\eps) =
\sum_{m=1}^{\infty} \frac{Z_{n,m}(v,c,g,u)}{\eps^m},
\label{eq:AZ}
\end{align}
where $Z_{n,m}(v,c,g,u)$ are finite functions of the couplings 
in the $\eps \rtarw 0$ limit.
The renormalized action is given by
the sum of the original action and the counter terms,
which can be expressed in terms of bare fields and bare couplings,
\begin{align}
\mathcal{S}_{ren} &= 
 \sum_{n=1}^4 
\sum_{\sigma=1}^{N_c} 
\sum_{j=1}^{N_f} 
\int \frac{d^{d+1}{k_B}}{(2\pi)^{d+1}} ~
\bar{\Psi}_{B;n,\sigma,j}(k_B) 
\Bigl[ i  \mathbf{\Gamma} \cdot \mathbf{K}_B  
+ i \gamma_{d-1} \veps_n( \vec k_B; v_B ) \Bigr] 
\Psi_{B;n,\sigma,j}(k)
\nn 
&  + \frac{1}{4} \int \frac{d^{d+1}q_B}{(2\pi)^{d+1}}
\Bigl[ \abs{\mathbf{Q}_B}^2 + ~ c_B^2 \abs{\vec q_B}^2
\Bigr] \tr{ \Phi_B(-q_B)  ~ \Phi_B(q_B) } \nn
  & + i \frac{g_B}{\sqrt{N_f}}
\sum_{n=1}^4
\sum_{\sigma,\sigma'=1}^{N_c}
\sum_{j=1}^{N_f} 
\int \frac{d^{d+1}{k_B}}{(2\pi)^{d+1}} 
\frac{d^{d+1}{q_B}}{(2\pi)^{d+1}} 
\Bigl[
\bar{\Psi}_{B;\bar n,\sigma,j}(k_B + q_B) 
\Phi_{B;\sigma,\sigma'}(q_B)
\gamma_{d-1} 
\Psi_{B;n,\sigma',j} (k_B) 
\Bigr]  \nn
&  + \frac{1}{4}
\int 
\frac{d^{d+1}{k}_{1B}}{(2\pi)^{d+1}} 
\frac{d^{d+1}{k}_{2B}}{(2\pi)^{d+1}} 
\frac{d^{d+1}{q}_B}{(2\pi)^{d+1}}
 \Bigl[ 
u_{1B} \tr{ \Phi_B(k_{1B} + q_B)  \Phi_B(k_{2B}-q_B) } \tr { \Phi_B(k_{1B})  \Phi_B(k_{2B}) } \nn
& +
u_{2B} \tr{ \Phi_B(k_{1B} + q_B)  \Phi_B(k_{2B}-q_B)   \Phi_B(k_{1B})  \Phi_B(k_{2B}) } 
 \Bigr].
\label{eq: S_ren_bare}
\end{align}
Here the renormalized quantities are related to the bare ones through
\begin{align}
 \mbf{K} & = \mc{Z}_\tau^{-1} ~ \mbf{K}_B, &
\vec k &= \vec{k}_B, \nn
 \Psi_{n,\sig,j}(k) &= \mc{Z}^{- \half}_\psi ~ \Psi_{B;n,\sig,j}(k_B), &
 \Phi(q) &= \mc{Z}^{- \half}_\phi ~ \Phi_B(q_B),  \nn
 v &= \frac{\mc{Z}_3}{\mc{Z}_2} ~ v_B, &
 c &=  \lt[\frac{\mc{Z}_\phi ~ \mc{Z}_\tau^{d-1}   }{\mc{Z}_5 }\rt]^{\half}  ~ c_B, \nn
 g &= \frac{\mc{Z}_\psi ~ \mc{Z}_\phi^\half  ~ \mc{Z}_{\tau}^{2(d-1)} }{\mc{Z}_6 } ~ \mu^{-\frac{3 - d}{2}} ~ g_B, &
 u_1 &= \frac{\mc{Z}_\phi^2 \mc{Z}_{\tau}^{3(d-1)} }{\mc{Z}_7 } ~ \mu^{-(3 - d)} ~ u_{1B}, \nn
 u_2 &= \frac{\mc{Z}_\phi^2 \mc{Z}_{\tau}^{3(d-1)} }{\mc{Z}_8 } ~ \mu^{-(3 - d)} ~ u_{2B},
\label{eq: bare_defn}
\end{align}
where
$\mc{Z}_\tau = \frac{\mc{Z}_1}{\mc{Z}_3}$, 
$\mc{Z}_\psi = \mc{Z}_1 ~\mc{Z}_{\tau}^{-d}$,
and
$\mc{Z}_\phi = \mc{Z}_4 ~\mc{Z}_{\tau}^{-(d +1)}$
with $ \mc{Z}_n = 1 + \mc{A}_n $.
We use the freedom of choosing an overall scale 
to fix the scaling dimension of $\vec k$ to be $1$.
The renormalized Green's function defined through
\begin{align}
& \langle \Psi(k_1) \ldots \Psi(k_f) \bar{\Psi}(k_{f+1})
\ldots \bar{\Psi}(k_{2f}) \Phi(q_1) \ldots \Phi(q_b) \rangle
\nn
& \qquad =  G^{(2f, b)}(k_i, q_j; v, c, g, u_i,  \mu) ~
\dl^{(d+1)} \lt(\sum_{i=1}^{f} (k_{i} - k_{i+f}) +
\sum_{j=1}^b q_{j} \rt)
\label{eq: rel_green_func} 
\end{align}
satisfies the renormalization group equation,
\begin{align}
& \lt[ z ~ \lt( \mbf{K}_i \cdot
\mbf{\nabla}_{\mbf{K}_i} + \mbf{Q}_j
\cdot \mbf{\nabla}_{\mbf{Q}_j} \rt) + \lt(\vec{k}_i \cdot
\vec{\nabla}_{\vec{k}_i} + \vec{q}_j
\cdot \vec{\nabla}_{\vec{q}_j} \rt) 
- \beta_v \frac{\dow}{\dow v} -  \beta_c \frac{\dow}{\dow c}
- \beta_g \frac{\dow}{\dow g} - \beta_{u_i} \frac{\dow}{\dow u_i}
\rt. \nn
& \quad \lt. + 2f \lt( \frac{d + 2}{2} -  \eta_\psi \rt) + b
\lt( \frac{d + 3}{2} -  \eta_\phi \rt) - \lt(z ~ (d-1) + 2
\rt) \rt] G^{(2f, b)}(k_i, q_i; v, c, g, u_i, \mu) = 0.
\label{eq: general_RG_eqn}
\end{align}
Here the dynamical critical exponent and the anomalous dimensions
of the fields are given by
$z   =   1 + \frac{\dow 
\ln{\mc{Z}_\tau}}{\dow \ln{\mu}}$,
$\eta_\psi   =
\half \frac{\dow \ln{\mc{Z}_\psi}}{\dow \ln{\mu}}$,
$\eta_\phi   = \frac{1}{2} 
\frac{\dow \ln{\mc{Z}_\phi}}{\dow \ln{\mu}}$,
and the beta functions that describe the flow of the
parameters with increasing energy scale are given by
$\beta_v  =  
\frac{\dow v}{\dow \ln{\mu}}$,
$\beta_c  =  
\frac{\dow c}{\dow \ln{\mu}}$,
$\beta_g  =  
\frac{\dow g}{\dow \ln{\mu}}$,
$\beta_{u_i}  =  
\frac{\dow u_i}{\dow \ln{\mu}}$. 
The set of coupled equations 
for the critical exponents and the beta functions
can be rewritten as
\begin{align}
 \mc{Z}_3 \lt[ (d-1)(z - 1) + 2\eta_\psi \rt] - \mc{Z}_3' 
&= 0, \nn
 \mc{Z}_1 \lt[ d(z - 1) + 2\eta_\psi \rt] - \mc{Z}_1' 
 &= 0, \nn
 \mc{Z}_4 \lt[ (d + 1)(z - 1) + 2\eta_\phi \rt] - \mc{Z}_4' 
 &= 0, \nn
 \mc{Z}_2 \lt[ \beta_v - v \lt\{(d - 1)(z - 1) + 2\eta_\psi 
\rt\} \rt] + v \mc{Z}_2' & = 0, \nn
 \mc{Z}_5 \lt[ 2 \beta_c - c \lt\{(d - 1)(z - 1) + 
2\eta_\phi \rt\} \rt] + c \mc{Z}_5' & = 0, \nn
 \mc{Z}_6 \lt[ \beta_g - g \lt\{- \frac{3-d}{2} + 2(d - 
1)(z - 1) + 2\eta_\psi + \eta_\phi \rt\} \rt] + g \mc{Z}_6'
& = 0, \nn
 \mc{Z}_7 \lt[ \beta_{u_1} - u_1 \lt\{-(3-d) + 3(d - 1)(z - 1) + 
4\eta_\phi \rt\} \rt] + u_1 \mc{Z}_7' & = 0, \nn
 \mc{Z}_8 \lt[ \beta_{u_2} - u_2 \lt\{-(3-d) + 3(d - 1)(z - 1) + 
4\eta_\phi \rt\} \rt] + u_2 \mc{Z}_8' & = 0, \nn
\label{eq: beta_simul}
\end{align}
which solve to give
\begin{align}
z &= \lt[ 1 + \lt( \half~ g \partial_g + u_i \partial_{u_i} \rt) 
( Z_{1,1} - Z_{3,1} ) \rt]^{-1}, 
\label{eq: z} \\
\eta_\psi &=  - \frac{\eps}{2}~ z  \lt( \half~ g \partial_g 
+ u_i \partial_{u_i} \rt) \lt( Z_{1,1} - Z_{3,1} \rt) + 
\frac{1}{2}~ z  \lt( \half~ g \partial_g + u_i \partial_{u_i} 
\rt) \lt( 2 Z_{1,1} - 3 Z_{3,1} \rt),  
\label{eq: eta_psi} \\
\eta_\phi &=  - \frac{\eps}{2}~ z  \lt( \half~ g \partial_g 
+ u_i \partial_{u_i} \rt) \lt( Z_{1,1} - Z_{3,1} \rt) + 
\frac{1}{2} ~ z \lt( \half~ g \partial_g + u_i \partial_{u_i} \rt) 
 \lt( 4 Z_{1,1} - 4 Z_{3,1} - Z_{4,1} \rt), 
\label{eq: eta_phi} \\
\beta_v &=    z ~ v \lt( \half~ g \partial_g + u_i \partial_{u_i} 
\rt) \lt( Z_{2,1} -  Z_{3,1} \rt), 
\label{eq: beta_v} \\
\beta_c &=   \frac{1}{2}~ z ~ c \lt( \half~ g \partial_g + u_i 
\partial_{u_i} \rt) \lt( 2Z_{1,1} -2 Z_{3,1} -  Z_{4,1} + 
Z_{5,1} \rt), 
\label{eq: beta_c} \\
\beta_g &=   - z ~g \lt[ \frac{\eps}{2} + \frac{1}{2}\lt( 
\half~ g \partial_g + u_i \partial_{u_i} \rt) \lt( 2 Z_{3,1} +  
Z_{4,1} - 2 Z_{6,1} \rt) \rt], 
\label{eq: beta_g} \\
\beta_{u_1} &=   - z ~ u_1 
\lt[ 
\eps -   
\lt( \half~ g \partial_g + u_i \partial_{u_i} \rt) 
\lt(  2 Z_{1,1} - 2  Z_{3,1} - 2  Z_{4,1} + Z_{7,1} \rt) 
\rt], 
\label{eq: beta_u1} \\
%
\beta_{u_2} &=   - z ~ u_2 
\lt[ 
\eps -   
\lt( \half~ g \partial_g + u_i \partial_{u_i} \rt) 
\lt(  2 Z_{1,1} - 2  Z_{3,1} - 2  Z_{4,1} + Z_{8,1} \rt) 
\rt].
\label{eq: beta_u2} 
\end{align}

\begin{figure}[!t]
\centering
\begin{subfigure}{0.3\textwidth}
 \includegraphics[width=\columnwidth]{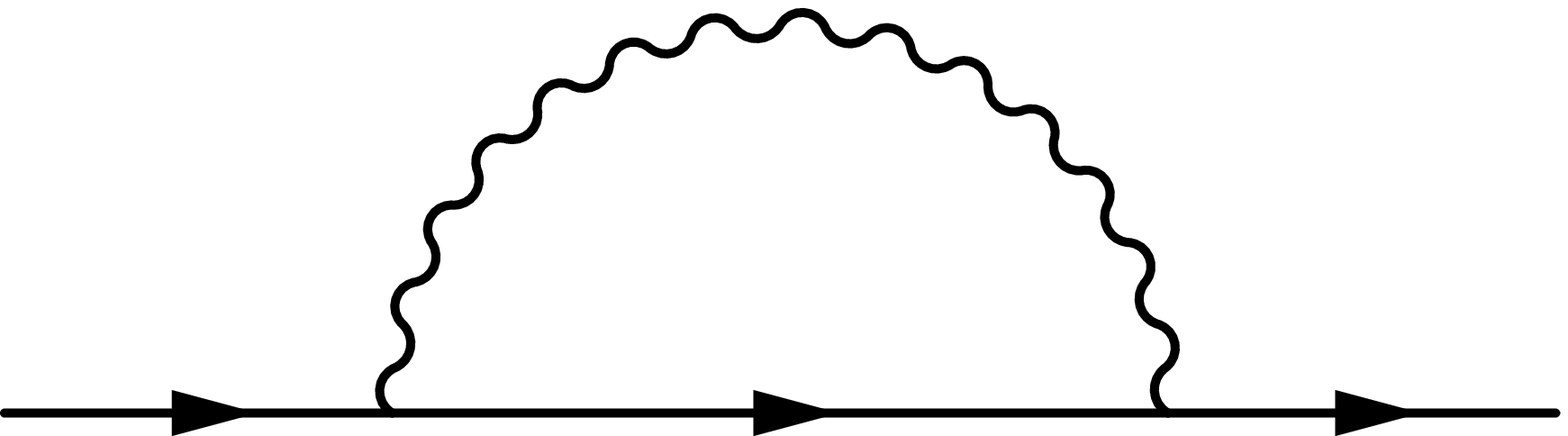}
 \caption{}
 \label{fig: CT_SEf}
\end{subfigure}
\quad
\begin{subfigure}{0.3\textwidth}
 \includegraphics[width=\columnwidth]{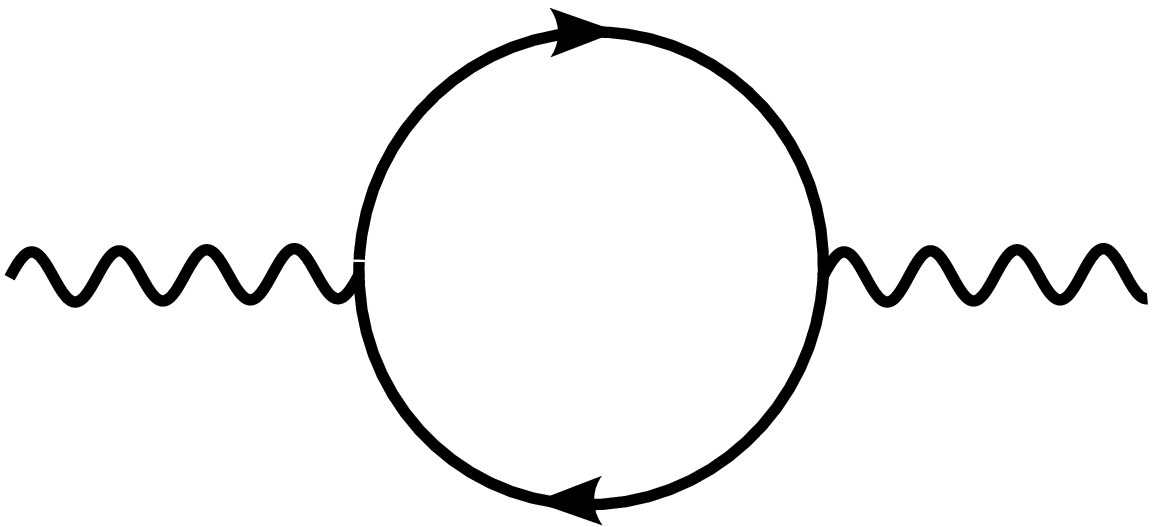}
 \caption{}
 \label{fig: CT_SEb}
\end{subfigure}
\quad
\begin{subfigure}{0.3\textwidth}
 \includegraphics[width=\columnwidth]{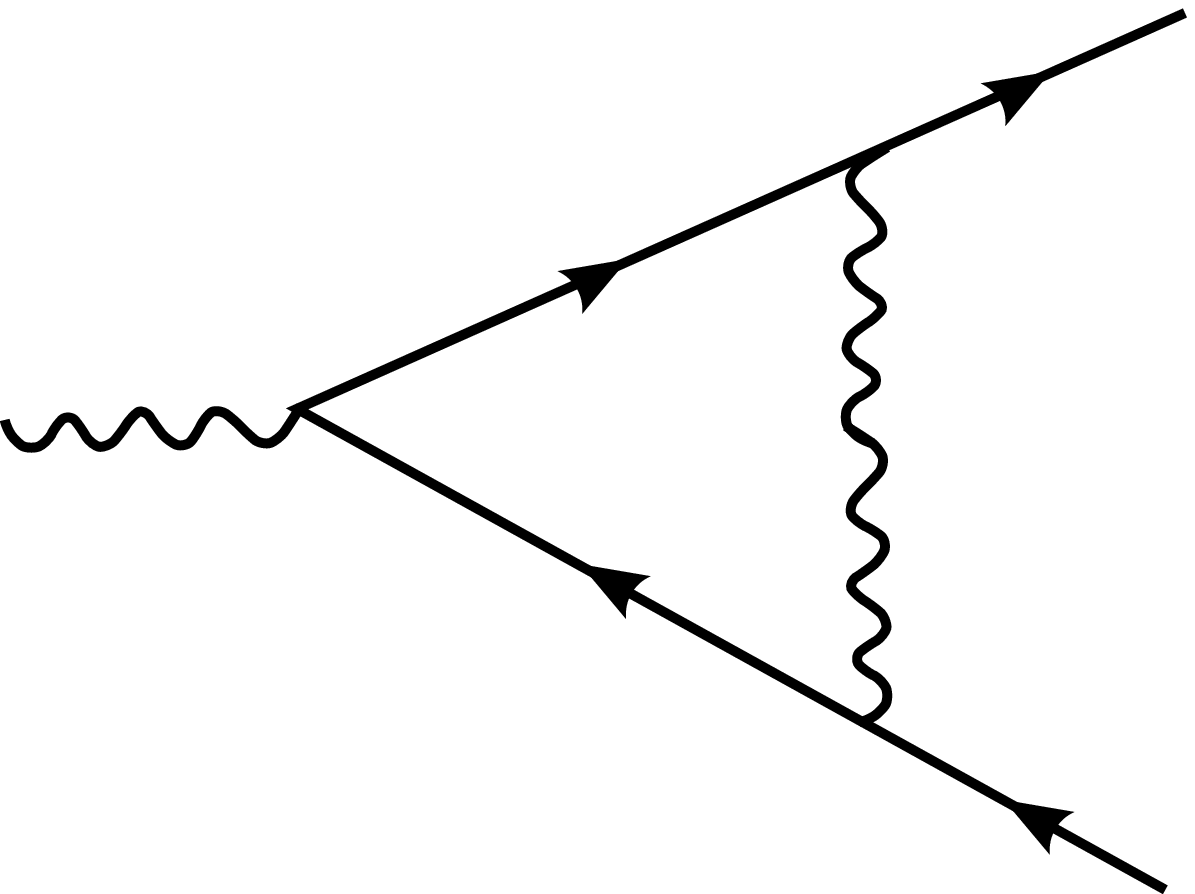}
 \caption{}
 \label{fig: CT_Y}
\end{subfigure} \\
\begin{subfigure}{0.22\textwidth}
 \includegraphics[width=\columnwidth]{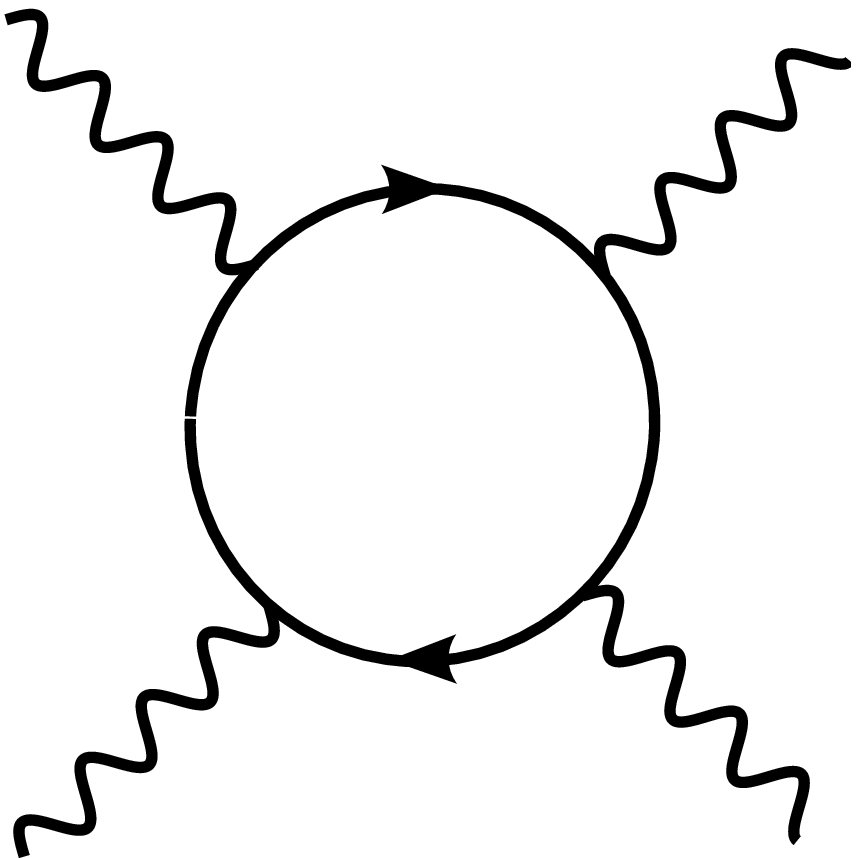}
 \caption{}
 \label{fig: CT_phi4-g}
\end{subfigure}
\quad
\begin{subfigure}{0.35\textwidth}
 \includegraphics[width=\columnwidth]{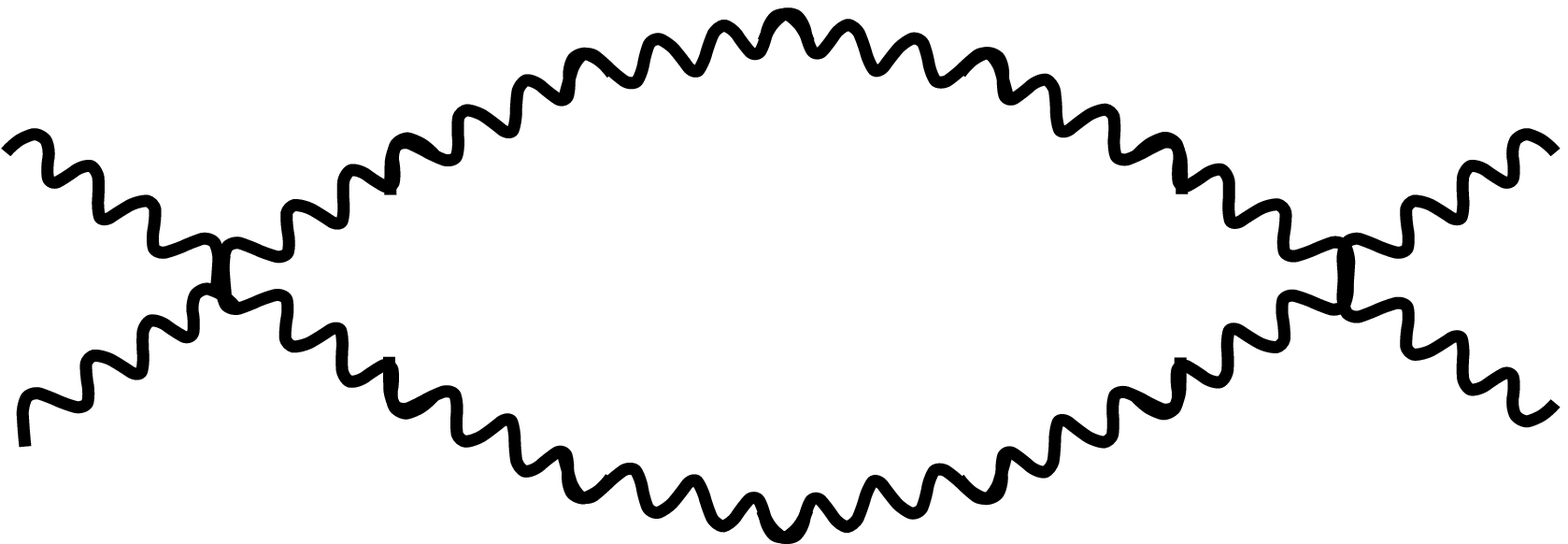}
 \caption{}
 \label{fig: CT_phi4-u}
\end{subfigure}
\caption{
One-loop Feynman diagrams that contribute
to the quantum effective action.
Solid (wiggly) lines represent the fermion (boson)
propagator.
Cubic vertices represent the Yukawa coupling, $g$.
In (e), each quartic vertex can be either $u_1$ and $u_2$.
}
\label{fig: CT}
\end{figure}

The counter terms can be computed order by order in the loop expansion.
We include the contributions from the one-loop diagrams
shown in Fig. \ref{fig: CT}.
The computations of the diagrams are discussed in the next section
of this supplementary material.
Here we summarize the final results,
\begin{align}
Z_{1,1} &=  - \frac{(N_c^2 - 1)}{4 \pi^2 N_c N_f } ~ 
\frac{g^2}{c} ~ h_1(v,c), \nn
Z_{2,1} &=   \frac{(N_c^2 - 1)}{4 \pi^2 N_c N_f } ~ 
\frac{g^2}{c} ~ h_2(v,c), \nn
Z_{3,1} &=  - Z_{2,1}, \nn
Z_{4,1} &=  - \frac{1}{4 \pi} ~ \frac{g^2}{v}, \nn
Z_{5,1} &=  0, \nn
Z_{6,1} &=  - \frac{1}{8\pi^3 N_c N_f} 
~ \frac{g^2}{c} ~ h_3(v,c), \nn
Z_{7,1} &= \frac{1}{2 \pi^2 c^2} \lt[
(N_c^2+7)u_1 
+ 2 \lt( 2 N_c - \frac{3}{N_c} \rt) u_2 
+ 3 \lt( 1 + \frac{3}{N_c^2} \rt) \frac{u_2^2}{u_1}  
\rt], \nn
Z_{8,1} &= \frac{1}{2 \pi^2 c^2} \lt[
12 u_1  
+ 2 \lt( N_c - \frac{9}{N_c} \rt) u_2  
\rt], 
\label{eq: 1L_Z_factors}
\end{align}
where $h_i(v,c)$ are defined in the main text.
This gives the beta functions 
and the dynamical critical exponent
shown in Eqs. (\ref{eq: beta_v_1L})-(\ref{eq: beta_u2_1L}) and below.
The anomalous dimensions of the fields are given by
\begin{align}
\eta_\psi &=   z ~ \frac{(N_c^2 - 1)}{8\pi^2 N_c N_f }~ 
\frac{g^2}{c} ~ \lt[  \eps ~\Bigl\{ h_1(v,c)  -
h_2(v,c) \Bigr\} -  \Bigl\{ 2 h_1(v,c) - 3 h_2(v,c)
\Bigr\} \rt] , 
\label{eq: eta_psi_1L} \\
\eta_\phi &= \frac{z}{8 \pi} \frac{g^2}{c} 
\lt[  \frac{ c}{v} - (4- \eps) 
\frac{(N_c^2-1)}{\pi N_c N_f} 
~\Bigl\{ h_1(v, c) - h_2(v, c) \Bigr\} \rt].
\label{eq: eta_phi_1L}
\end{align}
It is noted that the beta functions used in the main text
describe the flow of the couplings with increasing length scale,
which is defined to be
$\frac{\partial g}{\partial l} \equiv  - \beta_g$.

Because all $v,c,g,u_i$ flow to zero in the low energy limit
as discussed in the main text, 
it is more convenient to consider
the ratios of the couplings,
$w = \frac{v}{c}$, 
$\lambda = \frac{g^2}{v}$ and
$\kappa_i = \frac{u_i}{c^2}$.
The beta functions for the ratios are given by
\begin{align}
 \ParDeriv{w}{l} &=  \frac{z w \lambda}{8 \pi} 
\lt[ 1 - \frac{2w (N_c^2 -1)}{\pi N_c N_f} \Bigl\{ h_1(wc, 
c) + h_2(wc, c) \Bigr\} \rt], 
\label{eq: beta_w_1L} \\
\ParDeriv{\lambda}{l} &= z ~ \lambda  \lt[ \eps - 
\frac{\lambda}{4 \pi}  \lt\{ 1 - \frac{w}{\pi^2 N_c N_f}~ 
h_3(wc, w) \rt\} \rt],
\label{eq: beta_lambda_1L} \\
\ParDeriv{\kappa_1}{l} &= z ~ \kappa_1 \lt[ \lt\{ \eps - 
\frac{\lambda}{4\pi} \rt\} 
- \frac{(N_c^2 + 7)}{2 \pi^2} ~ \kappa_1
- \frac{(2 N_c^2 -3)}{\pi^2 N_c } ~ \kappa_2
- \frac{3(N_c^2 +3)}{2 \pi^2 N_c^2 } \frac{ \kappa_2^2}{\kappa_1}
 \rt],
\label{eq: beta_xi1_1L}  \\
\ParDeriv{\kappa_2}{l} &= z ~ \kappa_2 \lt[ \lt\{ \eps - 
\frac{\lambda}{4\pi} \rt\} 
- \frac{6}{\pi^2} ~ \kappa_1
- \frac{N_c^2 -9}{\pi^2 N_c } ~ \kappa_2
 \rt].
\label{eq: beta_xi2_1L} 
\end{align}
By using
$\lim_{c \rightarrow 0} h_1(wc,c)  =  \frac{\pi}{2}$, 
$\lim_{c \rightarrow 0} h_2(wc,c)  =  0$ and
$\lim_{c \rightarrow 0} h_3(wc,c)  =  \frac{2 \pi^2}{ 1 + w}$,
it can be shown that the beta functions for $c$, $w$, $\lambda$ and $\kappa_i$ 
simultaneously vanish at the attractive fixed point given in \eq{eq:fp}.
At the fixed point, 
the dynamical critical exponent and the anomalous dimensions
are given by
\bqa
z &=& 1+\frac{N_c^2+N_c N_f-1}{2(N_c^2+N_c N_f-3)} \epsilon, \nn
\eta_\psi  = \eta_\phi &=& -\frac{N_c^2+N_c N_f -1}{2(N_c^2+N_c N_f -3)} \epsilon
\eqa
to the leading order in $\epsilon$.
Both the dynamical critical exponent and the anomalous dimensions modify 
the scaling of the renormalized Green's function 
as can be checked from \eq{eq: general_RG_eqn}.
As a result, the two-point functions 
in Eqs.  (\ref{eq:G}) and (\ref{eq:D}) are controlled by
the net anomalous dimensions defined by
$\td \eta_\psi = \eta_\psi + \frac{(z-1)(2-\epsilon)}{2}$, 
$\td \eta_\phi = \eta_\phi + \frac{(z-1)(2-\epsilon)}{2}$,
which vanish to the linear order in $\epsilon$.
It is expected that there will be non-trivial anomalous dimensions 
for the two-point functions beyond the one-loop level\cite{PhysRevB.88.245106}.
Higher-point correlation functions exhibit 
non-trivial anomalous dimensions 
even to the linear order in $\epsilon$
because the quantum corrections are not 
canceled in \eq{eq: general_RG_eqn} for $2f+b>2$.

\section{Computation of one-loop diagrams} \label{app: 1Loop}

In this section, we outline
the computations of the one-loop
Feynman diagrams that result
in \eq{eq: 1L_Z_factors}.
We will use $\delta S$ to denote
the contributions to the quantum effective action,
and $S_{CT}$ to denote the counter terms
that are needed to cancel the UV divergent
pieces in $\delta S$
in the $\epsilon \rightarrow 0$ limit.

\subsection{Fermion self energy}

The quantum correction to the fermion self-energy
from the diagram in \fig{fig: CT_SEf} is
\begin{align}
\dl \mc{S}^{(2,0)} &= \frac{2 g^2 \mu^{3-d} }{N_f} \lt(N_c - 
\frac{1}{N_c}\rt)  
\sum_{n=1}^4 
\sum_{\sig=1}^{N_c} 
\sum_{j=1}^{N_f} 
\int \frac{d^{d+1} k}{(2\pi)^{d+1}} ~ 
\bar{\Psi}_{n,\sig,j}(k) 
~ \Upsilon_{(2,0)}^{(n)}(k) ~ \Psi_{n,\sig,j}(k), 
\label{eq: dS_20_defn}
\end{align}
where
\begin{align}
\Upsilon_{(2,0)}^{(n)}(k)
&= \int \frac{d^{d-1} \mbf{Q}}{(2 \pi)^{d-1}} \frac{d^2 
\vec q}{(2 \pi)^2}  ~ \gamma_{d-1} G_{\bar{n}}(k + q) 
\gamma_{d-1} ~ D(q),
\label{eq: upsilon_20_defn}
\end{align}
and the bare Green's functions are given by
\begin{align}
& G_n(k) = -i ~ \frac{\mathbf{\Gamma} \cdot \mbf{K} +
\gamma_{d-1} \veps_n(\vec k)}{\abs{\mbf{K}}^2 +
\veps_n^2(\vec k)}, \\
& D(q) = \frac{1}{\abs{\mbf{Q}}^2 + c^2 \abs{\vec q}^2}. 
\label{eq: propagators}
\end{align}
After the integrations over $\vec q$ and ${\bf Q}$,
\eq{eq: upsilon_20_defn} can be expressed in terms of
a Feynman parameter, 
\begin{align}
  \Upsilon_{(2,0)}^{(n)}(k)
&=  \frac{i}{ (4 \pi)^{(d+1)/2} c} 
\Gamma\lt(\frac{3-d}{2} \rt) 
\int_0^1 dx ~ \sqrt{\frac{1-x}{c^2 + x(1 + v^2 - c^2)}} \nn 
& \times \lt[ x(1-x) \lt\{ |\mbf{K}|^2 + \frac{c^2 
\veps_{\bar{n}}^2(\vec k)}{c^2 + x (1 + v^2 - c^2) } \rt\} 
\rt]^{-\frac{3-d}{2}} \lt[ \mbf{K} \cdot \mbf{\Gamma} - 
\frac{c^2 ~ \veps_{\bar{n}}(\vec k) ~ \gamma_{d-1}}{c^2 + x 
(1 + v^2 - c^2) }   \rt].
\end{align}
The UV divergent part in the $\eps \rightarrow 0$ limit is given by
$$\Upsilon_{(2,0)}^{(n)}(k)
=  \frac{i}{8 \pi^2 ~c ~  \eps} \lt[ h_{1}(v, c) ~
\mbf{K} \cdot  \mbf{\Gamma} - h_2(v, c) ~ 
\veps_{\bar{n}}(\vec k)\gamma_{d-1} \rt],$$  
where
$ h_1(v, c) = \int_0^1 dx ~ \sqrt{\frac{1-x}{c^2 + (1 + v^2
- c^2) x}}$, 
$ h_2(v, c) = c^2 \int_0^1 dx ~ \sqrt{\frac{1-x}{(c^2 + (1 +
v^2 - c^2) x)^3}}$. 
This leads to the one-loop counter term
for the fermion self-energy,
\begin{align}
 \mc{S}^{(2,0)}_{CT} 
 &= - i  \frac{g^2}{4\pi^2 c \eps} \frac{N_c^2 - 1}{N_f N_c}
\sum_{n=1}^4 
\sum_{\sig=1}^{N_c} 
\sum_{j=1}^{N_f} 
\int \frac{d^{d+1} k}{(2\pi)^{d+1}} ~ \nn
& \qquad \qquad \times
\bar{\Psi}_{n,\sig,j}(k) 
\lt[ h_{1}(v, c) ~
\mbf{K} \cdot  \mbf{\Gamma} - h_2(v, c) ~ 
\veps_{\bar{n}}(\vec k)\gamma_{d-1} \rt] 
\Psi_{n,\sig,j}(k).
\label{eq: CT_20}
\end{align}

\subsection{Boson self energy}

The boson self energy in \fig{fig: CT_SEb} is given by
\begin{align}
\dl \mc{S}^{(0,2)} &= - 2 g^2 \mu^{3-d} \sum_a \int \frac{d^{d+1} 
q}{(2 \pi)^{d+1}} ~ \Upsilon_{(0,2)}(q) ~ 
{\phi}^a(-q) {\phi}^a(q), \label{eq: dS_02_defn}
\end{align}
where
\begin{align}
\Upsilon_{(0,2)}(q) &=  \frac{1}{2} 
\sum_n
\int \frac{d^{d-1} 
\mbf{K}}{(2 \pi)^{d-1}} \frac{d^2 \vec k}{(2 \pi)^2} ~ 
\mbox{Tr}\lt[ \gamma_{d-1} G_{\bar n}(k + q) 
\gamma_{d-1} G_n(k) \rt].
\label{eq: upsilon_02_defn}
\end{align}
We first integrate over $\vec k$.
Using the Feynman parameterization, we write the resulting
expression as
\begin{align}
 \Upsilon_{(0,2)}(q) &=  \frac{1}{2 \pi v} 
 \int_0^1 dx \int \frac{d^{d-1} \mbf{K}}{(2 \pi)^{d-1}}
 ~ \frac{[x(1-x)]^{-\half} ~~ \mbf{K}\cdot (\mbf{K} + 
\mbf{Q})}
{x \abs{\mbf{K} + \mbf{Q}}^2 + (1 - x) \abs{\mbf{K}}^2}.
\end{align}
The quadratically divergent term is the mass renormalization,
which is automatically tuned away at the critical point in the present scheme.
The remaining correction to the kinetic energy of the boson becomes
$\Upsilon_{(0,2)}(q)  =  - \frac{\abs{\mbf{Q}}^2}{16 \pi v \eps} $
up to finite terms.
Accordingly we add the following counter term,
\begin{align}
 \mc{S}^{(0,2)}_{CT} &= - \sum_a \frac{g^2}{8 \pi v \eps}
 \int \frac{d^{d+1} q}{(2 \pi)^{d+1}} 
 ~ \abs{\mbf{Q}}^2 ~ {\phi}^a(-q) {\phi}^a(q).
\label{eq: CT_02}
\end{align}

\subsection{Yukawa vertex correction} \label{app: 
yukawa_1L}

The diagram in Fig. \ref{fig: CT_Y} gives rise to the vertex
correction in the quantum effective action,
\begin{align}
 \dl \mc{S}^{(2,1)} &= i \frac{2 g^3 \mu^{\frac{3}{2}(3-d) }}{N_c N_f^{3/2}}  
\sum_{a,n}
\sum_{j,\sigma,\sigma'} 
\int \frac{d^{d+1}{k}}{(2\pi)^{d+1}} 
\frac{d^{d+1}{q}}{(2\pi)^{d+1}} ~ 
{\phi}^a(q) 
\bar{\Psi}_{n,\sigma,j}(k+q) 
{\tau}^a_{\sigma,\sigma'} \Upsilon^{(n)}_{(2,1)}(k,q) ~ 
\Psi_{\bar n,\sigma',j}(k),
\label{eq: dS_21_defn}
\end{align}
where
\begin{align}
 \Upsilon^{(n)}_{(2,1)}(k,q) 
 &=  \int \frac{d^{d-1} \mbf{P}}{(2\pi)^{d-1}} 
\frac{d^{2} \vec p}{(2\pi)^{2}} ~ \gamma_{d-1} 
G_{\bar{n}}(p+q+k) \gamma_{d-1} G_{n}(p + k) 
\gamma_{d-1} ~ D(p).
\label{eq: upsilon_21_defn}
\end{align}
Here we use the identity for the $SU(N_c)$ generators,
$\sum_{a = 1}^{N_c^2 - 1}
\tau^{a} 
\tau^{b} 
\tau^{a} 
= - \frac{2}{N_c} ~ \tau^{b}$.
The UV divergent part in the $\epsilon \rightarrow 0$ limit,
which can be extracted by setting all external frequency and momenta 
to zero except $\mbf{K}$,
is given by
\begin{align}
 \Upsilon^{(n)}_{(2,1)}(\mbf{K}) 
 &=  \gamma_{d-1} \int \frac{d^{d-1} 
\mbf{P}}{(2\pi)^{d-1}} \frac{d^{2} \vec p}{(2\pi)^{2}} ~
 \frac{ | \mbf{P} |^2 - \veps_{\bar{n}}(\vec p) 
\veps_{n}(\vec p) }
{\Bigl[|\mbf{P}|^2 + c^2 |\vec p|^2 \Bigr] ~ 
\Bigl[ | \mbf{K} + \mbf{P} |^2  +  \veps_n^2(\vec p) \Bigr] 
~ \Bigl[ | \mbf{K} + \mbf{P} |^2  +  \veps_{\bar{n}}^2(\vec 
p) \Bigr]}.
\label{eq: upsilon_21_1}
\end{align}
We introduce two Feynman parameters to combine the denominators in 
the above expression.
In new coordinates $(R, \theta)$ defined by
$\veps_n(\vec p) = \sqrt{2 v} ~R \cos \theta$ and
$\veps_{\bar{n}}(\vec p) = \sqrt{2 v} ~R \sin \theta$,
\eq{eq: upsilon_21_1} is rewritten as
\begin{align}
 \Upsilon^{(n)}_{(2,1)}(\mbf{K}) 
 &= \gamma_{d-1} \frac{\Gamma(3)}{4 \pi^2} \int_0^1 dx_1 
\int_0^{1-x_1} dx_2 \int_0^{2 \pi} d\theta 
\int_0^{\infty} dR ~ R 
\int \frac{d^{d-1} \mbf{P}}{(2\pi)^{d-1}} ~ \nn
& \qquad \times \frac{| \mbf{P} |^2 - v~R^2 \sin(2\theta)}
{\Bigl[ | \mbf{P} |^2 + 2(x_1 + x_2) \mbf{K}\cdot \mbf{P} + 
M^2(v, c, x_1, x_2, \mbf{K}, R, \theta)  \Bigr]^3},
\label{eq: upsilon_21_3}
\end{align}
where
$M^2(v, c, x_1, x_2, \mbf{K}, R, \theta)  = (x_1 + x_2) 
\abs{ \mbf{K}}^2 + R^2  c ~ \zeta(v, c , x_1, x_2, 
\theta)$ 
with
$\zeta(v, c, x_1, x_2, \theta) 
 = \frac{2v}{c} ( x_1 \cos^2{\theta} + x_2 \sin^2{\theta} ) 
+ (1 - x_1 - x_2)\lt( vc  \cos^2(\theta + 
\pi/4) + \frac{c}{v}  \sin^2(\theta + \pi/4)\rt)$. 
Integrating over $\mbf{P}$ and $R$, 
we obtain 
$ \Upsilon^{(n)}_{(2,1)}(\mbf{K}) 
 =  \frac{\gamma_{d-1}}{16 \pi^3 ~ \eps ~ c}  ~ 
h_3(v, c) + \ordr{\eps^0}$ ,
where
$$ h_3(v, c) = \int_0^1 dx_1 \int_0^{1-x_1} 
dx_2 \int_0^{2\pi} d\theta \lt[ \frac{1}{\zeta(v, c, x_1, 
x_2, \theta)} -  \frac{v ~ \sin{2 
\theta}}{c ~ \zeta^2(v, c, x_1, x_2, \theta)}\rt].$$
It is noted that the UV divergent part 
of $\Upsilon^{(n)}_{(2,1)}$ is independent of $n$.
From this, we identify the counter term for the Yukawa vertex,
\begin{align}
  \mc{S}^{(2,1)}_{CT} &= - i  ~ 
\frac{g^3 \mu^{(3-d)/2} }{8 \pi^3 N_c N_f^{3/2} ~ \eps~c} ~ h_3(v, c)
\sum_{a,n}
\sum_{j,\sigma,\sigma'} 
\int \frac{d^{d+1}{k}}{(2\pi)^{d+1}} 
\frac{d^{d+1}{q}}{(2\pi)^{d+1}} \nn
& \quad \times {\phi}^a(q) \cdot 
\Bigl[\bar{\Psi}_{\bar n,\sigma,j}(k+q) 
{\tau}^a_{\sigma,\sigma'} \gamma_{d-1} \Psi_{n,\sigma',j} 
(k)  \Bigr].
\label{eq: CT_21}
\end{align}

\subsection{$\phi^4$ vertex corrections}
\label{sec:phi4}

There are two one-loop diagrams for the quartic vertex.   
The quantum correction from Fig. \ref{fig: CT_phi4-g} is
given by
\begin{align}
 \dl \mc{S}^{(0,4)}_{1} 
 &= \frac{1}{4} \frac{g^4 \mu^{2(3-d)} }{N_f^2} 
 \sum_{a_1,a_2,a_3,a_4 = 1}^{N_c^2 - 1} 
\int \frac{d^{d+1}q_1}{(2\pi)^{d+1}} 
\frac{d^{d+1}q_2}{(2\pi)^{d+1}} 
\frac{d^{d+1}q_3}{(2\pi)^{d+1}}
\frac{d^{d+1}q_4}{(2\pi)^{d+1}} 
\dl(q_1 + q_2 + q_3 + q_4) \nn
& \times  \Upsilon_{(0,4);1}(q_1,q_2,q_3) 
~ \tr{\tau^{a_1} \tau^{a_2} 
\tau^{a_3} \tau^{a_4}} ~ 
\phi^{a_1}(q_1) 
\phi^{a_2}(q_2) 
\phi^{a_3}(q_3) 
\phi^{a_4}(q_4),
\label{eq: dS_04-1_defn}
\end{align}
where 
\begin{align}
 & \Upsilon_{(0,4);1}(q_1,q_2,q_3) \nn
 &= \sum_n \int \frac{d^{d+1} k}{(2\pi)^{d+1}} 
\tr{\gamma_{d-1} G_n(q_1 + k) \gamma_{d-1} G_{\bar{n}}(q_1 
+ q_2 + k) \gamma_{d-1} G_n(q_1 + q_2 + q_3 + k) 
\gamma_{d-1} G_{\bar{n}}(k)}.
\label{eq: upsilon_04-1_defn}
\end{align}
When ${\bf Q}_i=0$, the above expression becomes
\begin{align}
 & \Upsilon_{(0,4);1}(q_1,q_2,q_3) 
= \sum_n \int \frac{d^{d+1} k}{(2\pi)^{d+1}} 
\mbox{tr} \Biggl[
\frac{1}{ \veps_n(\vec q_1 + \vec k) + \gamma_{d-1} {\bf K} \cdot {\bf \Gamma}}
\frac{1}{ \veps_{\bar n}(\vec q_1 +\vec q_2+ \vec k) + \gamma_{d-1} {\bf K} \cdot {\bf \Gamma}} \nn
& ~~~~~~~~~~~
\frac{1}{ \veps_n(\vec q_1 + \vec q_2 + \vec q_3 + \vec k) + \gamma_{d-1} {\bf K} \cdot {\bf \Gamma}}
\frac{1}{ \veps_{\bar n}(\vec k) + \gamma_{d-1} {\bf K} \cdot {\bf \Gamma}}
\Biggr].
\label{eq: upsilon_04-1_defn_3}
\end{align}
The matrix
$\gamma_{d-1} {\bf K} \cdot {\bf \Gamma}$
has an eigenvalue $i |{\bf K}|$ or $-i |{\bf K}|$.
Since the Green's functions in the trace
involve the common matrix, 
they always have poles on one side 
in the complex plane of $k_x$ ($k_y$)
for $n=1,3$ ($n=2,4$).
This is because $\veps_{n}(\vec k)$ and $\veps_{\bar n}(\vec k)$
have the same velocity in the $k_x$ ($k_y$) direction
for $n=1,3$ ($n=2,4$).
As a result, the integration over $\vec k$ vanishes
when the external ${\bf Q}_i$'s are zero.
Therefore no counter term is generated from  Fig. \ref{fig: CT_phi4-g}.

The diagram in Fig. \ref{fig: CT_phi4-u} represents
three different terms which are proportional to $u_1^2$,
$u_1 u_2$ and $u_2^2$.
It is straightforward to compute the counter terms
to obtain
\begin{align}
 \mc{S}^{(0,4)}_{CT} 
 &= \frac{\mu^{3-d}}{8 \pi^2 c^2 ~ \eps} 
\int \frac{d^{d+1} q_1}{(2\pi)^{d+1}}
\frac{d^{d+1} q_2}{(2\pi)^{d+1}} \frac{d^{d+1} 
q_3}{(2\pi)^{d+1}} \frac{d^{d+1} q_4}{(2\pi)^{d+1}} 
  ~ \dl(q_1 + q_2 + q_3 + q_4)  \nn
& 
\Bigg\{
\lt[ 
(N_c^2+7)u_1^2 
+ 2 \lt( 2N_c - \frac{3}{N_c} \rt) u_1 u_2
+ 3 \lt( 1 + \frac{3}{N_c^2} \rt) u_2^2
 \rt]
\tr{ \Phi(q_1) \Phi(q_2) } \tr{ \Phi(q_3) \Phi(q_4)} \nn
& +
\lt[ 
12 u_1 u_2  
+ 2 \lt(  N_c - \frac{9}{N_c} \rt) u_2^2
 \rt]
\tr{ \Phi(q_1) \Phi(q_2)  \Phi(q_3) \Phi(q_4)} 
\Bigg\}
.
\label{eq: CT_04}
\end{align}

\section{Beyond one-loop} 
\label{app: higherLoops}

The stability of the quasi-local strange metallic fixed point 
has been established to the one-loop order.
To examine higher-loop effects, 
one has to understand how general diagrams
depend on the couplings and velocities.
In the limit $v,c$ are small, the largest contributions 
come from the diagrams
where only nested hot spots are involved.
Therefore we focus on the diagrams which
have only $\Psi_{n,\sigma}$ and $\Psi_{\bar n, \sigma}$ for a fixed $n$.
Consider a general $L$-loop diagram 
that involves hot spots $n=1,3$
with $V_g$ Yukawa vertices
and $V_u$ quartic vertices,
\bqa
I \sim g^{V_g} u_i^{V_u} 
\int 
\left[
\prod_{i=1}^L d p_i 
\right]
\prod_{l=1}^{I_f} \left(
\frac{1}{ 
\mathbf{\Gamma} \cdot \mathbf{K}_l  
+  \gamma_{d-1} \left[ v k_{l,x} + (-1)^{\frac{n_l-1}{2}} k_{l,y} \right]   }
\right)
\prod_{m=1}^{I_b} \left(
\frac{1}{ 
| \mathbf{Q}_m |^2 + c^2 |\vec q_m|^2   }
\right). \nn
 \eqa 
Here both $u_1$ and $u_2$ are loosely denoted as $u_i$ 
because the power counting is equivalent for the two.
$k_l=({\bf K}_l, \vec k_l)$ and $q_m=({\bf Q}_m, \vec q_m)$ represent 
the momenta that go through
the fermion and boson propagators, respectively.
They are linear superpositions of
the internal momenta $p_i$ and external momenta.
$n_l$ is either $1$ or $3$.
Once $x$-components of all momenta are scaled by $1/v$, 
one has
\begin{align}
I &\sim \frac{ g^{V_g} u_i^{V_u} }{v^L} 
\int 
\lt[ \prod_{i=1}^L d p_i  \rt]
\prod_{l=1}^{I_f} \left(
\frac{1}{ 
\mathbf{\Gamma} \cdot \mathbf{K}_l  
+  \gamma_{d-1} \left[ k_{l,x} + (-1)^{\frac{n_l-1}{2}} k_{l,y} \right]   }
\right) \nn
& \qquad \qquad \qquad \quad \qquad \times 
\prod_{m=1}^{I_b} \left(
\frac{1}{ 
| \mathbf{Q}_m |^2 +  \frac{q_{m,x}^2}{w^2}  + c^2 q_{m,y}^2 )   }
\right). 
 \end{align}
The integrations of the internal momenta are well defined in the $v,c \rightarrow 0$
limit with fixed $w$  as far as each loop contains at least one fermion propagator.
The exceptions are the loops that are solely made of the boson propagators
for which the $y$-momentum integration is UV divergent for $c=0$.
The UV divergence is cut-off at $q_{m,y} \sim 1/c$ for each boson loops.
If there are $L_b$ boson loops, the entire diagram goes as
\bqa
I \sim \frac{ g^{V_g} u_i^{V_u} }{v^L c^{L_b}} = \lambda^{\frac{V_g+2-E}{2}} \kappa_i^{V_u} w^{-V_u} g^{(E-2)} c^\delta,
\label{eq:exp}
\eqa
where $E$ is the number of external lines
and $\delta = V_u - L_b \geq 0$.
Here we used the identity $L = \frac{V_g + 2 V_u + 2 - E}{2}$.

\begin{figure}[!t]
\centering
\begin{subfigure}{0.3\textwidth}
 \includegraphics[width=\columnwidth]{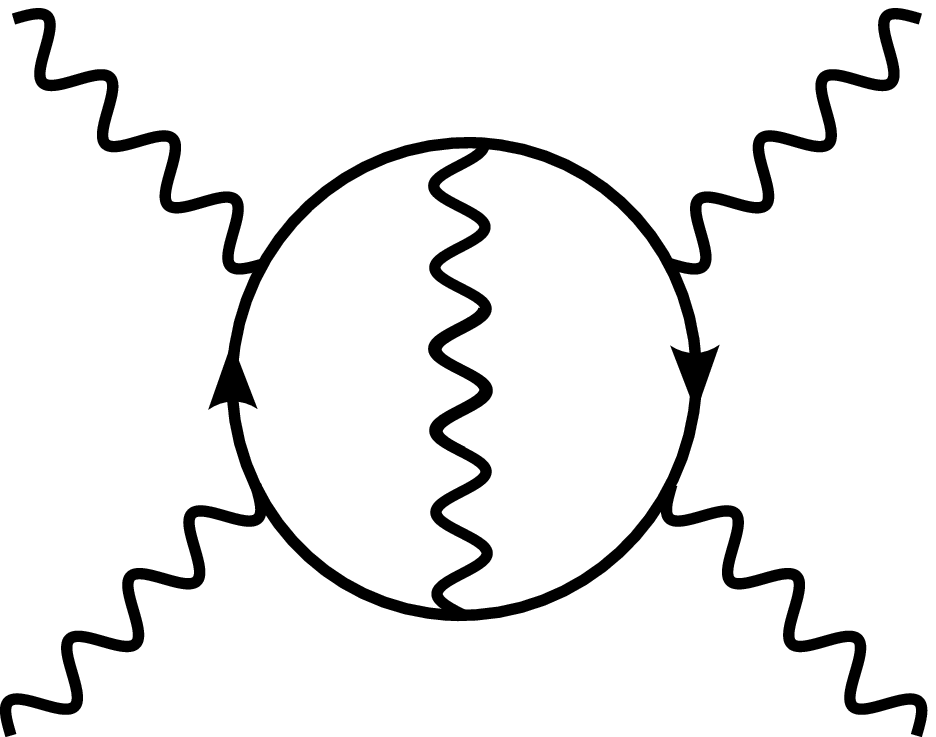}
\caption{}
 \label{fig: phi4_2L-1}    
\end{subfigure}
\qquad
\begin{subfigure}{0.28\textwidth}
 \includegraphics[width=\columnwidth]{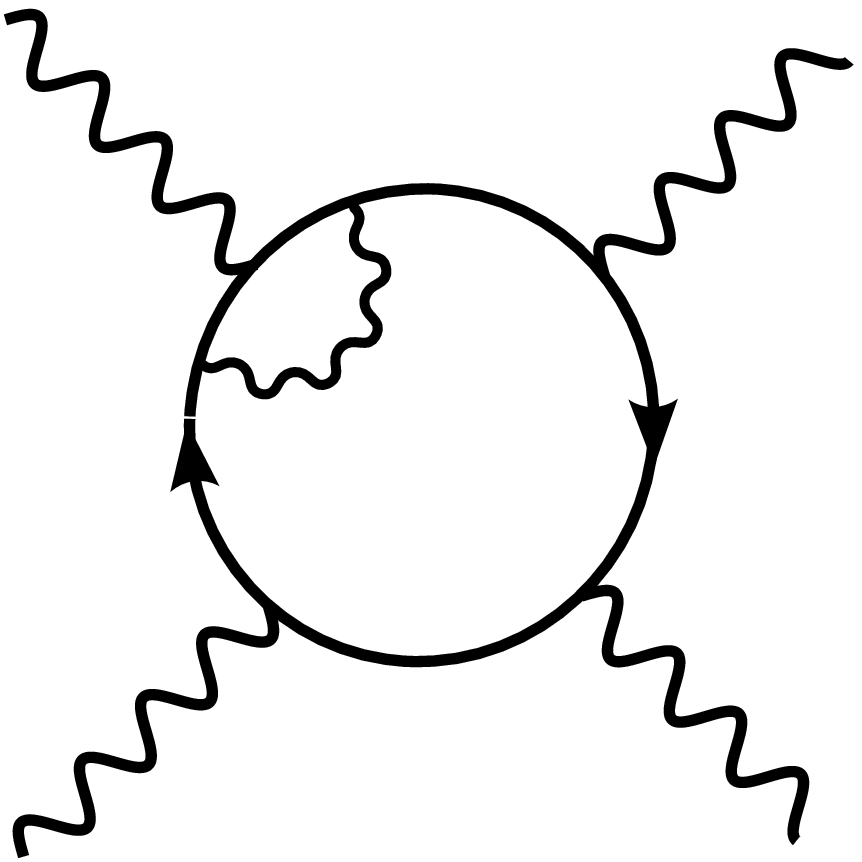}
\caption{}
 \label{fig: phi4_2L-2}    
\end{subfigure}
\caption{Non-vanishing two-loop corrections to the $\phi^4$
vertex 
that do not contain $u_1$ or $u_2$.}
\label{fig: phi4_2L}
\end{figure}

\eq{eq:exp} implies that the multiplicative renormalizations 
for the kinetic energy ($E=2$), the Yukawa coupling ($E=3$) and
the quartic vertices ($E=4$)
defined in \eq{eq:AZ} go as
\bqa
\mathcal{A}_1, .., \mathcal{A}_5 & \sim & \lambda^{\frac{V_g}{2}} ~ \kappa_i^{V_u} ~ w^{-V_u} ~ c^\delta, \nn
\mathcal{A}_6         & \sim & \lambda^{\frac{V_g-1}{2}}~  \kappa_i^{V_u} ~ w^{-V_u}~  c^\delta, \nn
\mathcal{A}_7 & \sim & \lambda^{\frac{V_g-2}{2}} ~ \kappa_i^{V_u} ~ w^{-V_u} ~ c^\delta ~ \frac{g^2}{u_7}, \nn
\mathcal{A}_8 & \sim & \lambda^{\frac{V_g-2}{2}}~  \kappa_i^{V_u}~  w^{-V_u}~  c^\delta~  \frac{g^2}{u_8}.
\eqa
$\mathcal{A}_1, .., \mathcal{A}_6$ remain finite 
in the limit $v,c,g,u_i$ go to zero with fixed $\lambda, w, \kappa_i$.
Therefore, higher-loop corrections in $\mathcal{A}_1, .., \mathcal{A}_6$ are systematically
suppressed by powers of $\lambda$, $\kappa_i$.
On the other hand, $\mathcal{A}_7$ and $\mathcal{A}_8$ are proportional to
$\lambda^{\frac{V_g}{2}} ~ \kappa_i^{V_u-1} ~ w^{-V_u+1} ~ c^{\delta-1}$.
Since the higher-loop quantum corrections with $\delta \geq 1$ are obviously suppressed, 
we will focus on the contributions
with $\delta=0$ 
in $\mathcal{A}_7$ and $\mathcal{A}_8$ 
which go as $1/c$ in the $c \rightarrow 0$ limit.
Only diagrams with $\delta =0$ are the ones 
that do not contain $u_i$.
At the one-loop order, there is one such diagram for the $\phi^4$ vertices,
 Fig. \ref{fig: CT_phi4-g}.
Due to a chiral structure that is present in the one-loop diagram,
it vanishes as is shown in Sec. \ref{sec:phi4}.
Higher-loop diagrams with $\delta =0$ do not vanish in general.
For example, 
the two-loop diagrams in Fig. \ref{fig: phi4_2L}
generate quantum corrections for $\kappa_i$
which are order of $\frac{\lambda^3}{c}$.
At $d=3$, these higher-loop corrections 
are still vanishingly small in the low energy limit.
This is because $\lambda$ vanishes as $1/l$ 
while $c$ vanishes only as $1/log(l)$ in the $l \rightarrow \infty$ limit,
where $l$ is the logarithmic length scale.
Since all higher-loop corrections are suppressed at $d=3$,
the one-loop beta functions become asymptotically exact 
in the low energy limit where $\lambda$, $\kappa_i$ vanish
along with $v$, $c$.
For $d<3$, the higher-loop quantum corrections to $\kappa_i$
grow as $c$ becomes small with $\lambda \neq 0$.
This suggests that $c$ should be stabilized at a nonzero value once higher-loop corrections are included.
In particular, 
$\kappa_i$ enters into the beta function of $c$
at the three-loop and higher orders.
It is expected that the feedback of $\kappa_i$ will  
stabilize $c$ at a nonzero value in the low energy limit.
Once the velocity becomes nonzero, 
$\kappa_i$ will flow to a nonzero and finite value at the fixed point.
Because of the continuity from the exact $d=3$ fixed point,
not only $c, v, \kappa_i, \lambda$ 
but also $\frac{\lambda^n}{c}$ with $n \geq 3$
at the fixed point in $d=3-\epsilon$
should go to zero in the $\epsilon \rightarrow 0$ limit.
Therefore, higher order corrections 
including the corrections to $\kappa_i$ with $\delta=0$
are systematically suppressed,
and the expansion is controlled for small $\epsilon$.

\section{Enhancement of superconducting and charge density wave fluctuations} \label{app: suscep}

In this section, we compute the anomalous dimensions
of the superconducting (SC) and charge density wave (CDW) operators
that are enhanced at the strange metallic fixed point.

\subsection{Anomalous dimension} \label{app: 
rho_scaling}

We consider an insertion of a fermion bilinear, 
\begin{align}
 \mc{S}_{\rho} &= \rho ~\mu 
 \int \frac{d^{d+1} k}{(2\pi)^{d+1}} 
\wtil \Psi_{n,\sig,j}(k) 
~ \Omega_{n,\sig,j; n^{'},\sig^{'},j^{'}} 
~\Psi_{n^{'},\sig^{'},j^{'}}(k).
\label{eq: schematic_insertion}
\end{align}
Here $\rho$ is a dimensionless source.
$\wtil \Psi_{n,\sig,j}(k)$ is either $\bar \Psi_{n,\sig,j}(k)$ or $\Psi_{n,\sig,j}^T(-k)$
depending on whether the operator creates particle-hole or particle-particle excitations.
$\Omega_{n,\sig,j; n^{'},\sig^{'},j^{'}}$
is a matrix that specifies 
the momentum, spin and flavor quantum numbers of the insertion.
The UV divergence in the quantum effective action coming from 
the insertion is canceled by a counter term of the same form,
$\mc{S}_{\rho; CT} = \rho ~ \mu ~ \mc{A}_\rho \int 
\frac{d^{d+1} k}{(2\pi)^{d+1}}  ~ 
\wtil \Psi_{n,\sig,j}(k) 
~ \Omega_{n,\sig,j; n^{'},\sig^{'},j^{'}} 
~\Psi_{n^{'},\sig^{'},j^{'}}(k)$
with $ \mc{A}_\rho = \sum_{ m} \frac{Z_{\rho,m}}{\eps^m}$.
The renormalized insertion can be written as
\begin{align}
 \mc{S}_{\rho; ren} &= \rho_B 
 \int \frac{d^{d+1} k_B}{(2\pi)^{d+1}}~ 
\wtil \Psi_{B; n,\sig,j}(k) 
~ \Omega_{n,\sig,j; n^{'},\sig^{'},j^{'}} 
~\Psi_{B; n^{'},\sig^{'},j^{'}}(k),
\label{eq: schematic_insertion_ren}
\end{align}
where the renormalized source $\rho$ is related to the bare 
source $\rho_B$ as 
$ \rho = \mu^{-1} ~ \frac{\mc{Z}_\psi  ~ \mc{Z}_{\tau}^{d-1}}{\mc{Z}_\rho} ~ \rho_B$
with 
$\mc{Z}_\rho = 1 + \mc{A}_\rho$.
From this, one can obtain the beta function for the source,
\begin{align}
 \frac{d \rho}{d l} &=  \rho \lt[ 1 + \gamma_\rho \rt],
\label{eq: beta_rho_1L}
\end{align}
where 
$\gamma_{\rho} =  
z \lt( \frac{ g}{2} ~\partial_g + u_i ~\partial_{u_i} \rt) ( Z^{}_{3,1} - Z^{}_{\rho,1} )$
is the anomalous dimension of the source.
The larger the anomalous dimension of the source is, 
the stronger the enhancement is.

\subsection{Superconducting channel} \label{app: pairing}
\begin{figure}[!t]
\centering
\begin{subfigure}{0.4\textwidth}
 \includegraphics[width=\columnwidth]{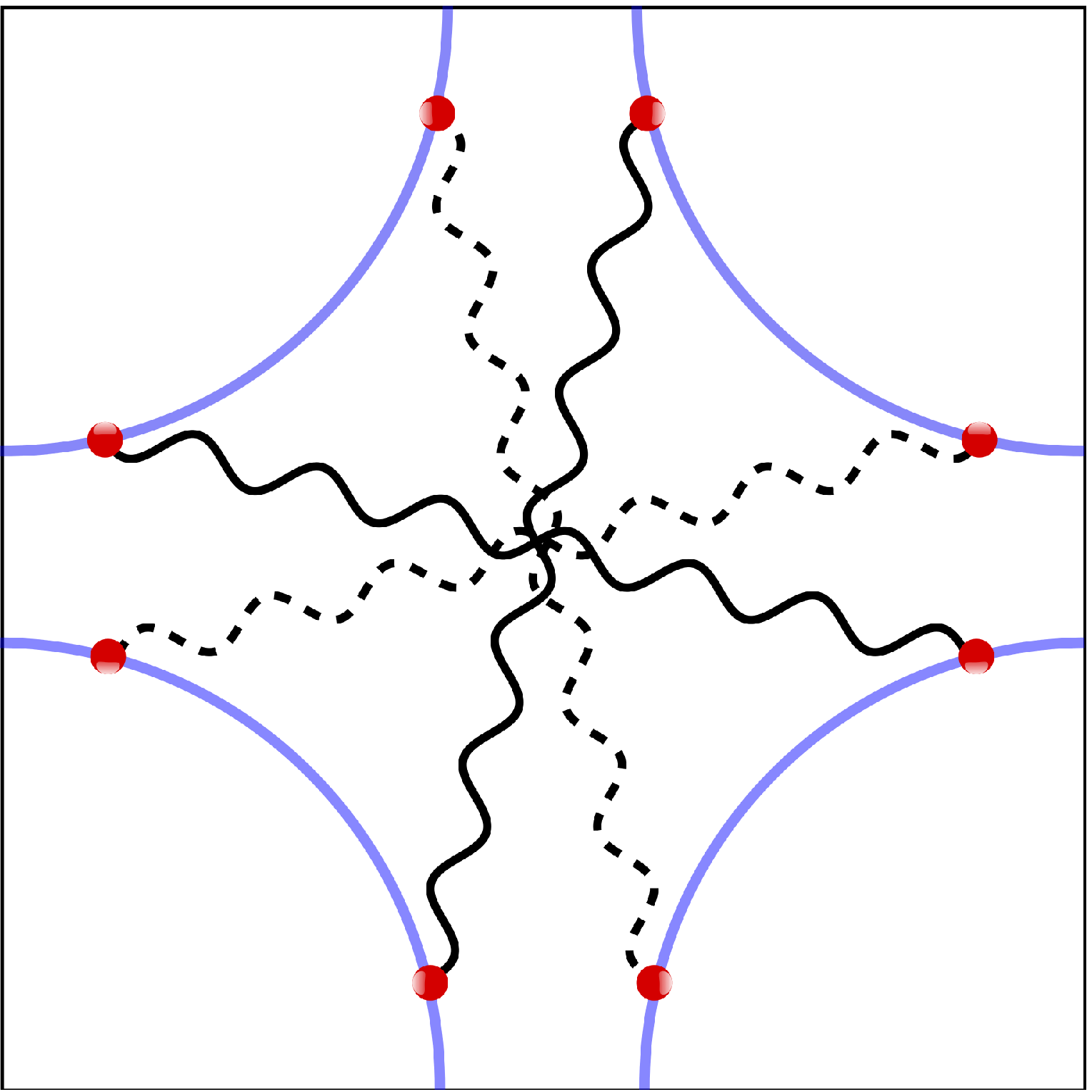}
 \caption{~$g$-wave}
 \label{fig: SC_2-2}
\end{subfigure}
\qquad
\begin{subfigure}{0.4\textwidth}
 \includegraphics[width=\columnwidth]{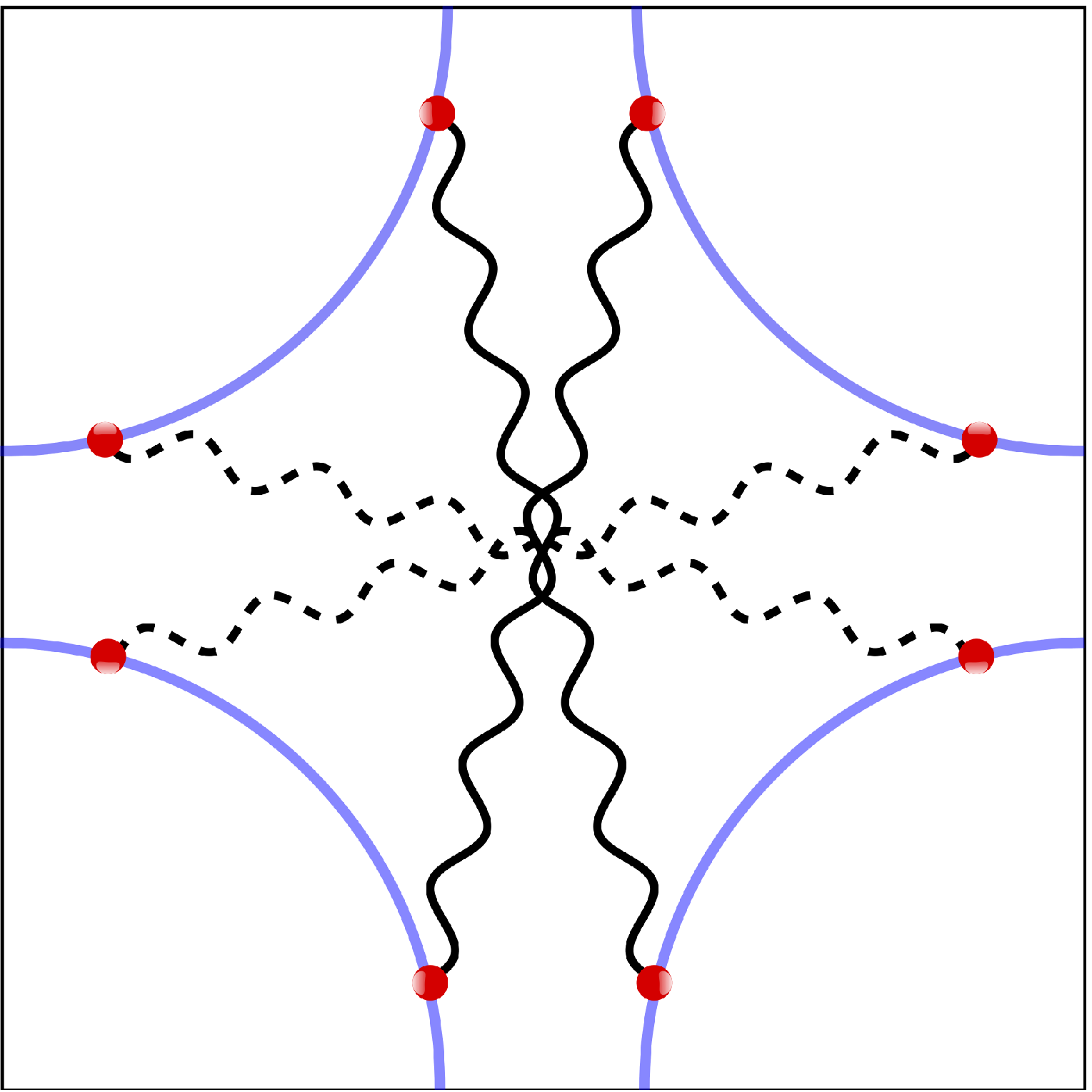}
 \caption{~$d_{x^2 - y^2}$-wave}
 \label{fig: SC_2-1}
\end{subfigure}
\caption{
Cooper pair wavefunctions represented 
by the vertices
(a) $\mc{S}^{( + )}_{A,\gamma_{d-1}}$ and
(b) $\mc{S}^{( - )}_{A,\gamma_{d-1}}$.
A wiggly line connecting two momenta $k_1$ and $k_2$
represents a Cooper pair made of electrons at those momenta.
The dashed wiggly lines are intended to represent 
the relative minus sign in the Cooper pair wavefunction 
relative to the ones connected by the solid lines.
The Cooper pair created by $\mc{S}^{( + )}_{A,\gamma_{d-1}}$  
( $\mc{S}^{( - )}_{A,\gamma_{d-1}}$ ) undergoes four (two) phase winding
under $2\pi$ rotation, which correspond to $g$ ($d_{x^2-y^2}$) wave pairing.
} 
\label{fig: SC_zero}
\end{figure}

\begin{figure}[!t]
\centering
\begin{subfigure}{0.4\textwidth}
 \includegraphics[width=\columnwidth]{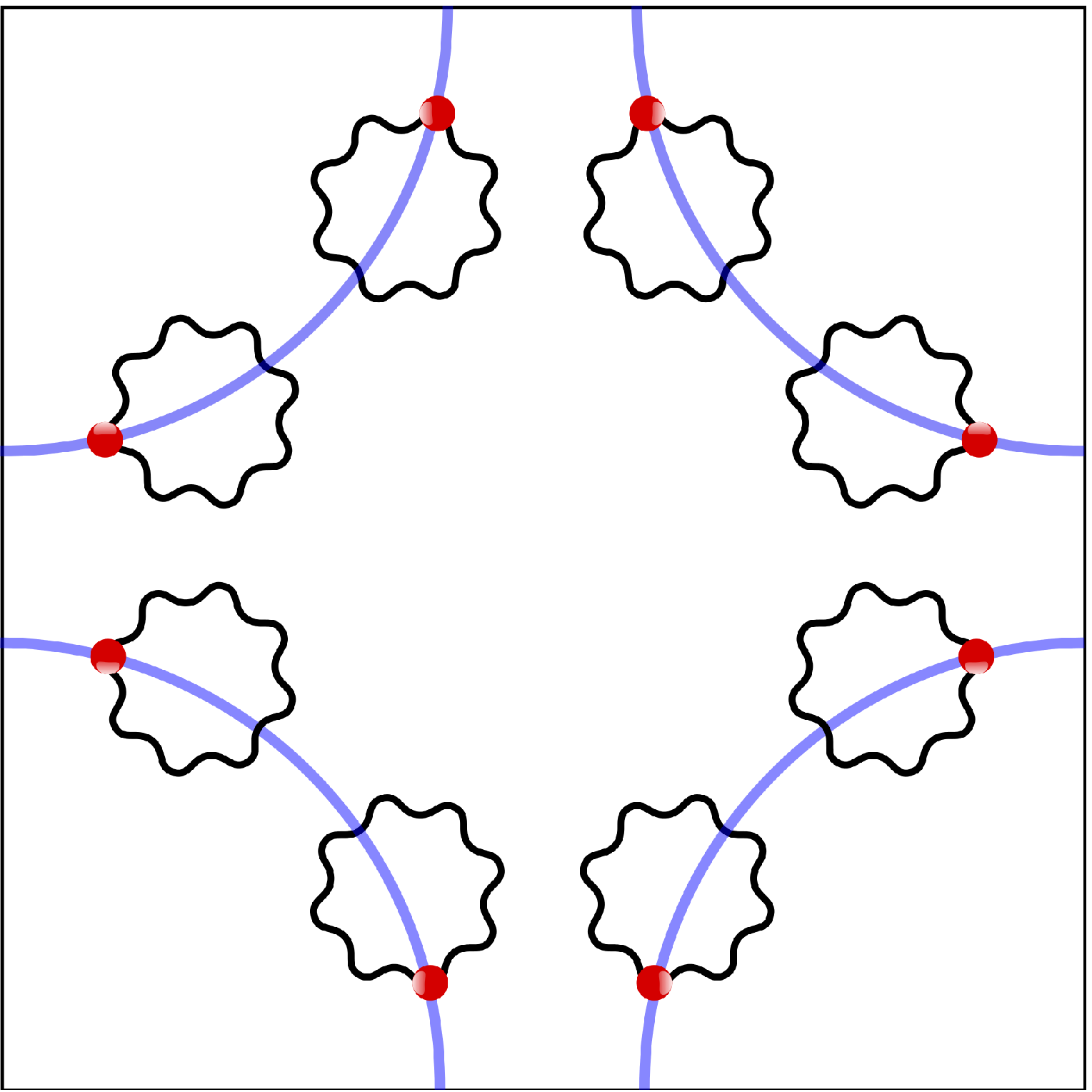}
 \caption{~$s$-wave}
 \label{fig: SC_5-1}
\end{subfigure}
\qquad
\begin{subfigure}{0.4\textwidth}
 \includegraphics[width=\columnwidth]{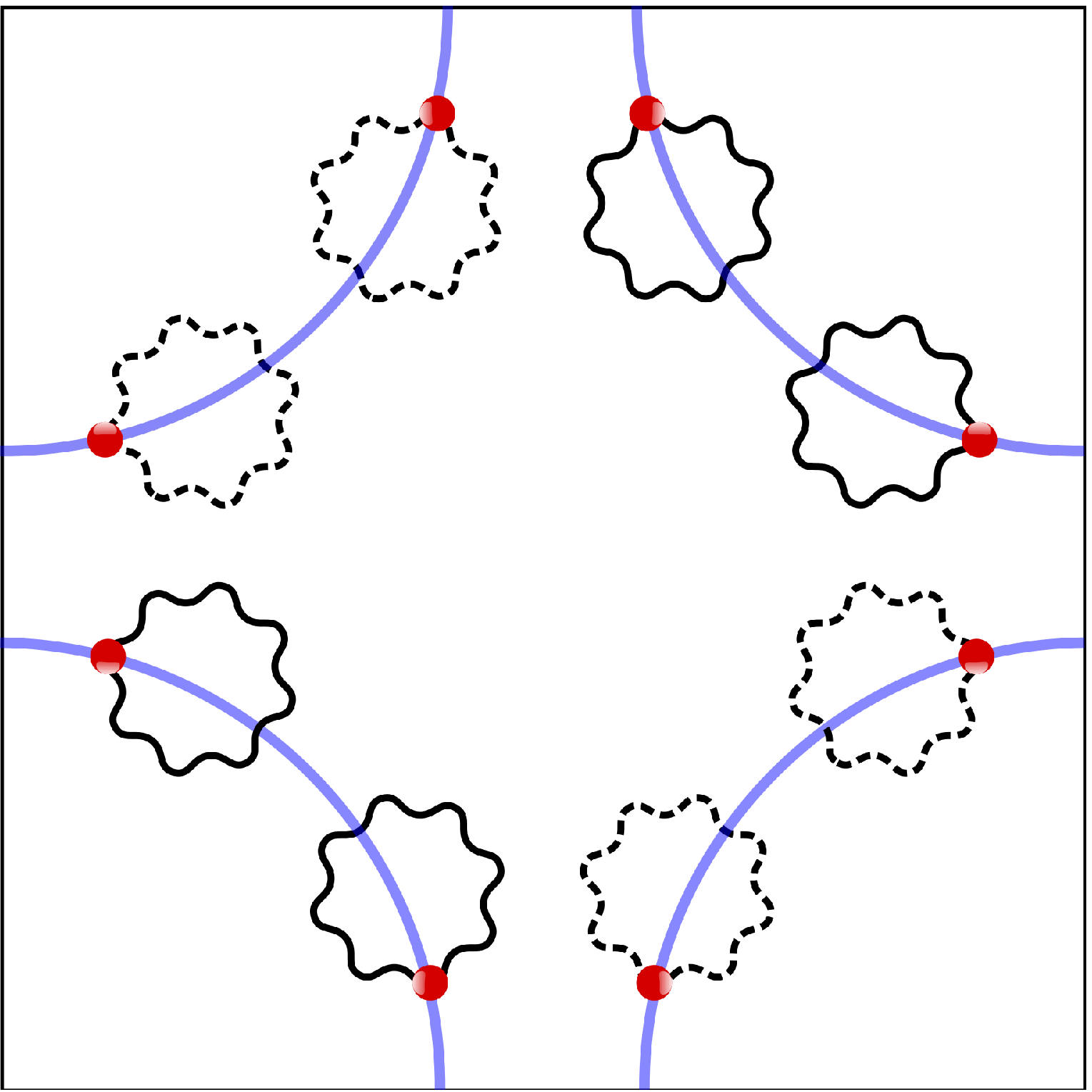}
 \caption{~$d_{xy}$-wave}
 \label{fig: SC_5-2}
\end{subfigure}
\caption{
Cooper pair wavefunctions represented 
by the vertices 
(a) $\mc{S}^{( + )}_{A,I}$ and
(b) $\mc{S}^{( - )}_{A,I}$.
A wiggly line that ends on a hot spot represents
a Cooper pair made of electrons from that hot spot.
Therefore, the Cooper pair carries non-zero momenta, $2 \vec
k_F$.
The Cooper pair created by $\mc{S}^{( + )}_{A,I}$  
( $\mc{S}^{( - )}_{A,I}$ ) undergoes 
zero (two) phase winding
under $2\pi$ rotation, 
which correspond to $s$ ($d_{xy}$) wave pairing.
}
\label{fig: SC_2Kf}
\end{figure}

Here we examine the superconducting channels 
described by the pairing vertices of the form,
\begin{align}
\mc{S}^{( \pm )}_{A,\what{\Omega}} &=  \mu V \sum_{j,\sig,\sig'} 
\int \frac{d^{d+1} k}{(2 \pi)^{d+1}} \lt[ \lt\{
\Psi_{1,\sig,j}^{T}(-k) A_{\sig,\sig'}~ \what{\Omega}~ 
\Psi_{1,\sig',j}(k) + \Psi_{3,\sig,j}^{T}(-k) A_{\sig,\sig'} 
~\what{\Omega} ~\Psi_{3,\sig',j}(k) \rt\} \rt. \nn
& \qquad \qquad \qquad \lt. \pm \lt\{
\Psi_{2,\sig,j}^{T}(-k) A_{\sig,\sig'} ~\what{\Omega} ~
\Psi_{2,\sig',j}(k) 
+ 
\Psi_{4,\sig,j}^{T}(-k) A_{\sig,\sig'} ~\what{\Omega} ~ 
\Psi_{4,\sig',j}(k)
\rt\} \rt].
\label{eq: pairing_action}
\end{align}
Here $V$ is a source for the pairing operator.
$A_{\sig,\sig'}$ is an anti-symmetric matrix,
which represents the spin-singlet pairing for the case of $N_c=2$.
$\what{\Omega}$ is a $2 \times 2$ matrix that acts on the Dirac indices.
Among all possible $\what{\Omega}$, 
we find the channels with $\what{\Omega}=\gamma_{d-1}$ and $I$
are most strongly enhanced.
Therefore we will focus on these channels in the rest of the section.
$\mc{S}^{( + )}_{A,\gamma_{d-1}}$ ( $\mc{S}^{( - )}_{A,\gamma_{d-1}}$ )
describes the $g$-wave ($d_{x^2-y^2}$-wave) pairing with zero net momentum of Cooper pairs
as is illustrated in Fig. \ref{fig: SC_zero}.
$\mc{S}^{( + )}_{A,I}$ ( $\mc{S}^{( - )}_{A,I}$ )
describes the Cooper pairs
with non-zero net momentum $2 \vec k_F$ 
in the $s$-wave ($d_{xy}$-wave) channel
as is shown in 
Fig. \ref{fig: SC_2Kf}.

The one-loop quantum correction to the SC insertion is given by
\begin{align}
\dl \mc{S}^{( \pm )}_{A,\what{\Omega}} &=    N_{V} \mu^{4-d} V ~
g^2 \sum_{j,\sig,\sig'} \int \frac{d^{d+1} k}{(2 \pi)^{d+1}} \nn
& \quad \times \lt[
\lt\{ 
\Psi_{1,\sig,j}^T(-k) A_{\sig,\sig'}
\Upsilon_{\what{\Omega}}^{(1)}(k) \Psi_{1,\sig',j}(k) 
+ \Psi_{3,\sig,j}^T(-k) A_{\sig,\sig'} \Upsilon_{\what{\Omega}}^{(3)}(k) 
\Psi_{3,\sig',j}(k)  \rt\} \rt. \nn
& \qquad  \qquad \lt. \pm \lt\{
\Psi_{2,\sig,j}^T(-k) 
A_{\sig,\sig'}\Upsilon_{\what{\Omega}}^{(2)}(k) 
\Psi_{2,\sig',j}(k) 
+ \Psi_{4,\sig,j}^T(-k)
A_{\sig,\sig'} \Upsilon_{\what{\Omega}}^{(4)}(k)
\Psi_{4,\sig',j}(k) \rt\} \rt],
\label{eq: d_SC_defn}
\end{align}
where
\begin{align}
\Upsilon_{\what{\Omega}}^{(n)}(k) &= \int \frac{d^{d+1} q}{(2
\pi)^{d+1}} ~ D(q) ~ \gamma_{d-1}^{T} G_{\bar n}^{T}(-k-q)
~ \what{\Omega} ~ G_{\bar n}(k+q) \gamma_{d-1}
\label{eq: upsilon_SC_defn}
\end{align}
and
$N_V = \frac{2(N_c+1)}{N_c N_f}$.
Using $\gamma_0^T  =  - \gamma_0$ and
$\gamma_{i}^T = \gamma_{i}$, for $i = 1,2, \ldots, (d-1)$,
we obtain 
\begin{align}
& \Upsilon_{\what{\Omega}}^{(n)}(\mbf{K})   = \int \frac{d^{d+1} q}{(2 \pi)^{d+1}} \nn
& \frac{ \lt[ (K_0 + Q_0) \gamma_0 - \sum_{\nu =1}^{d-2}(K_\nu + Q_\nu) 
\gamma_\nu + \veps_{\bar n}(\vec q) \gamma_{d-1}\rt] \gamma_{d-1} 
~\what{\Omega}~ \gamma_{d-1}  \lt[ - (\mbf{K} + 
\mbf{Q} ) \cdot \mbf{\Gamma} + \veps_{\bar n}(\vec q) \gamma_{d-1} 
\rt]}
{ \lt[\abs{\mbf{Q}}^2 + c^2 \abs{\vec q}^2 \rt] ~ 
\lt[ \abs{\mbf{K} + \mbf{Q}}^2 +  \veps_{\bar n}^2(\vec q) \rt]^2
}
\label{eq: upsilon_SC_2}
\end{align}
when $\vec k=0$.
Changing coordinates from $(q_x, q_y)$ to $(R, \theta)$ with
$\veps_{\bar n}(\vec q) = \sqrt{2v} R \cos{\theta}$ and
$\veps_{n}(\vec q) = \sqrt{2v} R \sin{\theta}$,
one can perform the integrations over $R$ and ${\bf Q}$ using 
the Feynman parameterization to obtain
$$\Upsilon_{\what{\Omega}}^{(n)}(\mbf{K}) = 
\frac{1}{16 \pi^3 c ~ \eps} ~ \what{\Omega} ~ h_{SC}(v, c)
+ \ordr{\eps^0},$$
where
$h_{SC}(v, c) = \frac{2v}{c} \int_0^1 dx \int_0^{2 \pi} d\theta ~ 
\frac{x ~ \cos^2{\theta}}{ \zeta_{1}^2(v, c, x, \theta)}
$
with
$$\zeta_{1}(v, c, x, \theta) = \frac{2 v x}{c}
\cos^2{\theta} + (1 - x)\lt[ \frac{c}{v} \sin^2\lt( \theta
+ \frac{\pi}{4} \rt)   + v c \cos^2\lt( \theta
+ \frac{\pi}{4} \rt) \rt].$$
Note that $h_{SC}(v, c)$ is same for $\what{\Omega} = \gamma_{d-1}, I$.
Therefore, we add the counter term, 
\begin{align}
\mc{S}_{A,\what{\Omega}; CT}^{(\pm)} 
&=  -  \mu V ~ \frac{N_V}{16 \pi^3  \eps}  ~
\frac{g^2 }{c} h_{SC}(v, c) \sum_{\sig,\sig'} \int
\frac{d^{d+1} k}{(2 \pi)^{d+1}} \nn
& \quad \times \lt[ \lt\{
\Psi_{1,\sigma}^{T}(-k) A_{\sig,\sig'} ~\what{\Omega}~ 
\Psi_{1,\sigma'}(k) + \Psi_{3,\sigma}^{T}(-k) A_{\sig,\sig'} 
~\what{\Omega}~ \Psi_{3,\sigma'}(k) \rt\} \rt. \nn
& \qquad \qquad \qquad \lt. \pm \lt\{
\Psi_{2,\sigma}^{T}(-k) A_{\sig,\sig'} ~\what{\Omega}~ 
\Psi_{2,\sigma'}(k)
+ \Psi_{4,\sigma}^{T}(-k) A_{\sig,\sig'} ~\what{\Omega}~
\Psi_{4,\sigma'}(k) \rt\} \rt],
\label{eq: CT_SC}
\end{align}
which gives the anomalous dimension of the source,
$ \gamma_{V}  =  
\frac{N_V}{8 \pi^2} 
\frac{z ~ g^2 }{c} 
\lt[ \frac{1}{2 \pi} h_{SC}(v, c) - (N_c-1) h_2(v,c) \rt]$
for the four vertices,
$\mc{S}^{( \pm )}_{A,\gamma_{d-1}}$ and $\mc{S}^{( \pm )}_{A,I}$.
At the quasi-local strange metal fixed point, we have
$\lim_{c \rightarrow 0} h_2(w^* c, c) = 0$ and 
$\lim_{c \rightarrow 0} h_{SC}(w^* c,c) = \pi^2$,
and the anomalous dimension becomes
$ \gamma_{V}  =   \frac{\lambda^*}{8\pi  (N_c-1)}$.

It is interesting that 
the finite momentum pairing 
is as strong as the zero momentum 
pairing. 
This is a consequence of the nesting,
which allows a pair of electrons to stay on the Fermi surface
as they are scattered from one hot spot to another.
The attractive interaction 
is mediated by the commensurate spin fluctuations
which scatter a pair of electrons in one hot spot 
to another hot spot, e.g., from $1+$ to $1-$.
Because one electron in the Cooper pair 
with momentum $2 \vec k_F$
is above the Fermi surface
and the other is below the Fermi surface,
the matrix element for the scattering is negative
at low frequencies.
As a result, the interaction is attractive
in the symmetric combination.

\subsection{Charge density wave channel} \label{app: DW} 
\begin{figure}[!t]
\centering
\begin{subfigure}{0.4\textwidth}
 \includegraphics[width=\columnwidth]{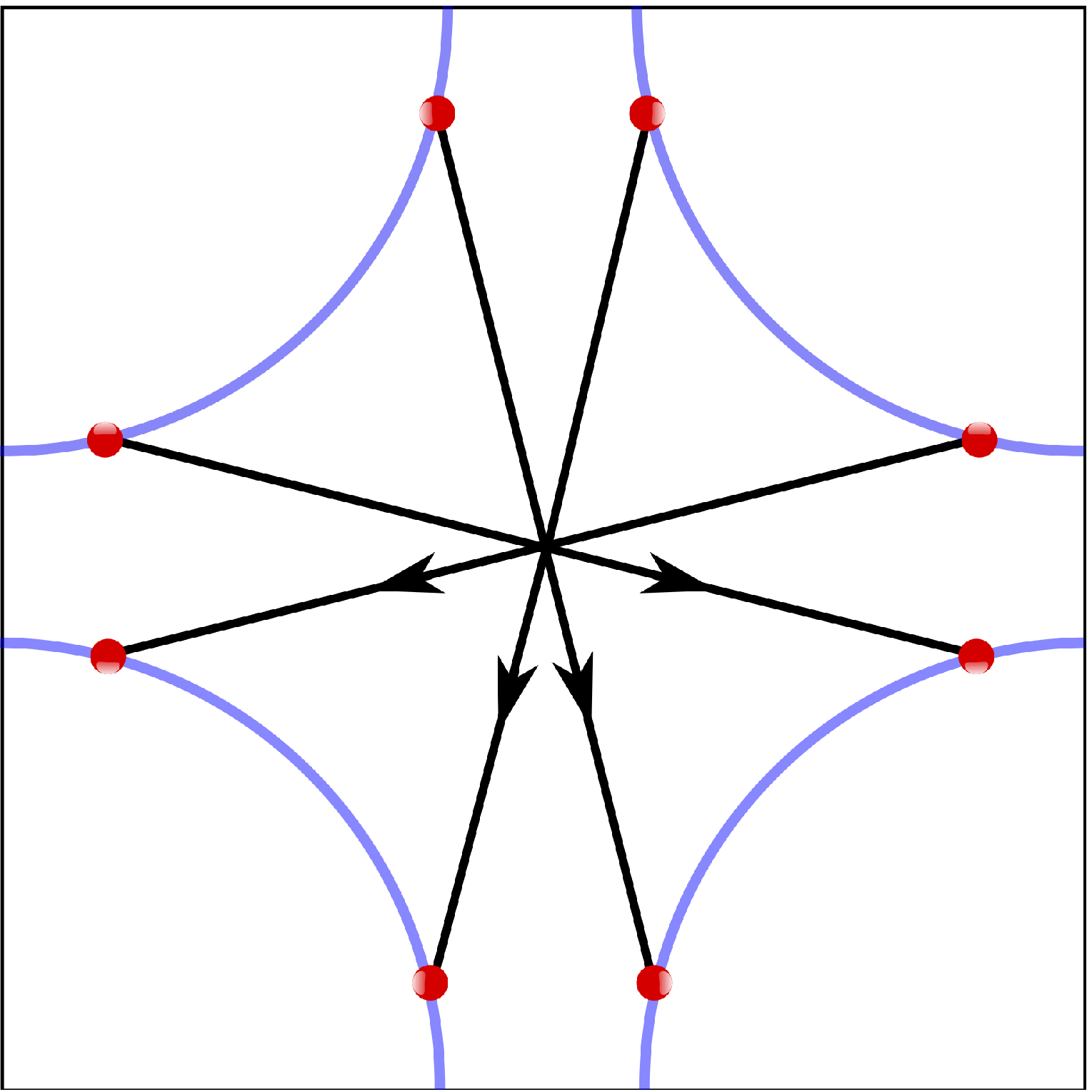}
\caption{~$p_y$-wave}
 \label{fig: CDW_3-1}    
\end{subfigure}
\qquad
\begin{subfigure}{0.4\textwidth}
 \includegraphics[width=\columnwidth]{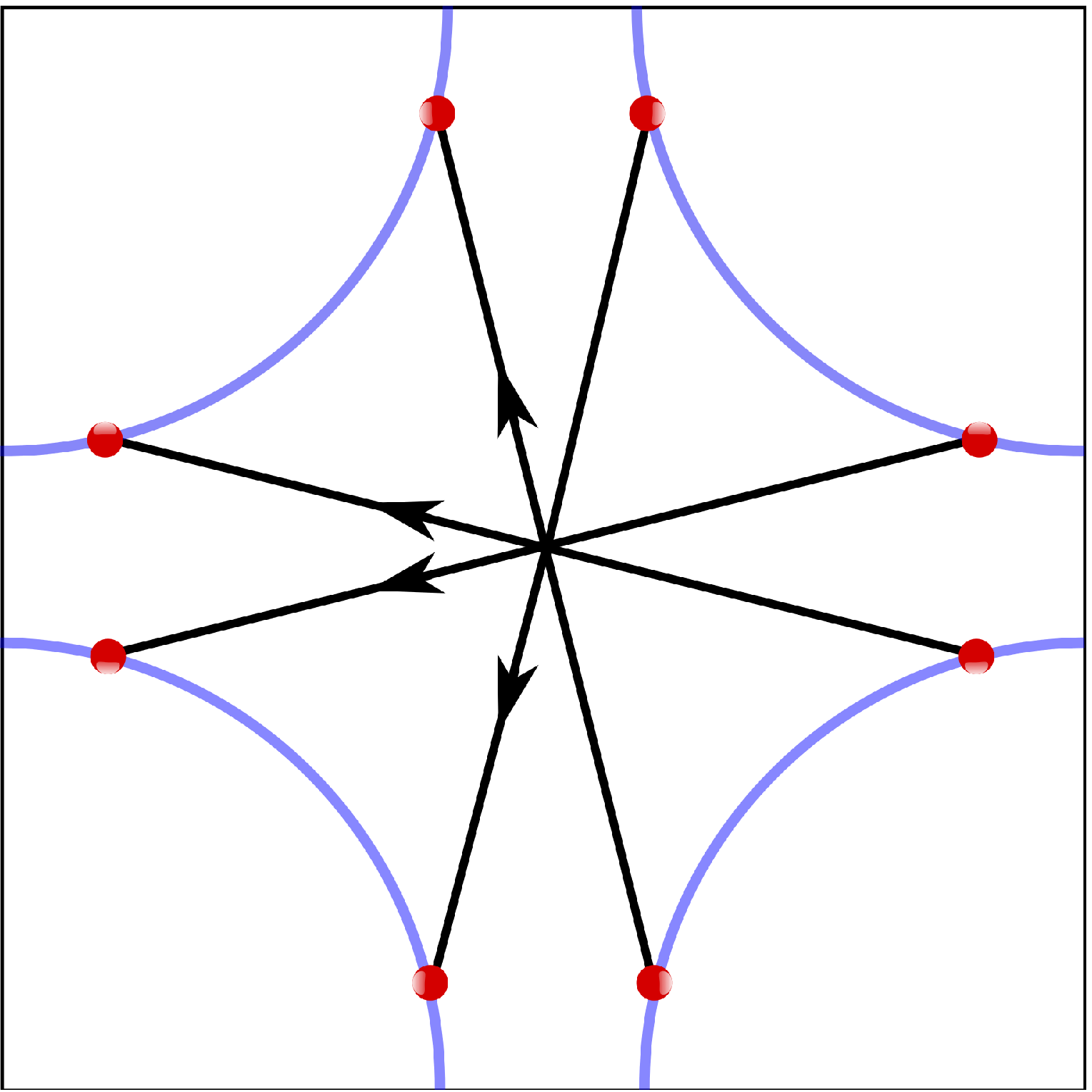}
\caption{~$p_x$-wave}
 \label{fig: CDW_3-2}    
\end{subfigure}
\caption{
Wavefunctions of the particle-hole pairs
created by the vertices 
(a) $\mc{S}^{( + )}_{\rho}$ and
(b) $\mc{S}^{( - )}_{\rho}$.
An arrow from $k_2$ to  $k_1$ represents
a particle-hole pair created by 
$i ( c_{k_1}^* c_{k_2} - c_{k_2}^* c_{k_1} )$,
where $c_k$ is the electron field at momentum $k$ 
with spin and flavor indices suppressed.
$\mc{S}^{( + )}_{\rho}$ 
($\mc{S}^{( - )}_{\rho}$)
is odd under the $y$ ($x$) reflection,
while both of them preserve time-reversal.
} 
\label{fig: 2kf_CDW}
\end{figure}

Here we compute anomalous dimensions for CDW operators of the form,
\begin{align}
\mc{S}_{\rho}^{(\pm)} &=  \rho^{(\pm)} \mu \sum_{j,\sig}
\int \frac{d^{d+1} k }{(2 \pi)^{d+1}} \nn
& \quad \times 
\lt[ \lt\{ \bar{\Psi}_{1,\sig,j}(k) \Psi_{1,\sig,j}(k) 
+ \bar{\Psi}_{3,\sig,j}(k) \Psi_{3,\sig,j}(k) \rt\} 
\pm \lt\{ \bar{\Psi}_{4,\sig,j}(k) \Psi_{4,\sig,j}(k) 
+ \bar{\Psi}_{2,\sig,j}(k) \Psi_{2,\sig,j}(k) \rt\} \rt].
\label{eq: rho4_action}
\end{align}
$\mc{S}_{\rho}^{(+)}$ ($\mc{S}_{\rho}^{(-)}$)
describes $p_y$-wave ($p_x$-wave) CDW 
which carries momentum $2\vec k_F$
as is shown in Fig. \ref{fig: 2kf_CDW}.
These operators are pseudospin singlets for $N_c=2$ 
and has no SC counterpart connected by the pseudospin transformation.
The one-loop quantum correction is given by
\begin{align}
\dl \mc{S}_{\rho}^{(\pm)} &=  - N_\rho ~ \rho^{(\pm)}  \mu^{4-d}
g^2 \sum_{j,\sig}
\int \frac{d^{d+1} k }{(2 \pi)^{d+1}} 
\lt[ \lt\{ \bar{\Psi}_{1,\sig,j}(k) 
\Upsilon_{\rho}^{(1)}(k) \Psi_{1,\sig,j}(k) 
+ \bar{\Psi}_{3,\sig,j}(k) \Upsilon_{\rho}^{(3)}(k) 
\Psi_{3,\sig,j}(k) \rt\} \rt. \nn
& \qquad \qquad \qquad  \lt. \pm \lt\{ 
\bar{\Psi}_{2,\sig,j}(k) 
\Upsilon_{\rho}^{(2)}(k) \Psi_{2,\sig,j}(k) 
+ \bar{\Psi}_{4,\sig,j}(k) \Upsilon_{\rho}^{(4)}(k) 
\Psi_{4,\sig,j}(k) \rt\} \rt],
\label{eq: dS_rho4_defn}     
\end{align}
where
\begin{align}
\Upsilon_{\rho}^{(n)}(k) &= \int \frac{d^{d+1} q
}{(2 \pi)^{d+1}} ~ D(q) ~ \gamma_{d-1} G_{\bar n}(k+q) ~ 
G_{\bar n}(k+q) \gamma_{d-1}
\label{eq: upsilon_rho4_defn}
\end{align}
and $N_{\rho} = \frac{2}{N_f} \lt( N_c - \frac{1}{N_c} \rt)$.
From a straightforward calculation, we identify the counter term
\begin{align}
\mc{S}_{\rho; CT}^{(\pm)} 
&= - \rho^{(\pm)} \mu ~ \frac{N_{\rho}}{16 \pi^3 \eps} ~ 
\frac{g^2}{c} ~ h_{CDW}(v, c) \sum_{j,\sig}
\int \frac{d^{d+1} k }{(2 \pi)^{d+1}} 
\lt[ \lt\{\bar{\Psi}_{1,\sig,j}(k) \Psi_{1,\sig,j}(k) +
\bar{\Psi}_{3,\sig,j}(k) \Psi_{3,\sig,j}(k) \rt\} \rt. \nn
& \qquad \qquad \qquad \lt. \pm \lt\{ 
\bar{\Psi}_{4,\sig,j}(k)
\Psi_{4,\sig,j}(k) + \bar{\Psi}_{2,\sig,j}(k)  
\Psi_{2,\sig,j}(k) \rt\} \rt],
\label{eq: CT_rho1}
\end{align}
where
$ h_{CDW}(v, c) = \int_0^1 dx \int_0^{2\pi} d\theta ~ 
x \lt[ \frac{1}{\zeta_{1}(v, c, x, \theta)} + \frac{2v}{c}~
\frac{\cos^2{\theta}}{\zeta^2_{1}(v, c, x, \theta)} \rt]$.
From this we find the anomalous dimension of the CDW source, 
$ \gamma_{CDW}^{(\pm)} = 
\frac{N_\rho}{16 \pi^3} \frac{z ~ g^2 }{c} \Bigl[ h_{CDW}(v, c) - 2\pi ~ h_2(v,c) 
\Bigr] $.
At the fixed point, we have 
$\lim_{c \rightarrow 0} h_{CDW}(w^* c, c) = 2\pi^2$
and the anomalous dimension becomes $\gamma_{CDW}^{(\pm)} = \frac{\lambda^*}{4\pi}$.


\input{ref.tex}

\end{document}